\documentclass{aa}
\usepackage{amsmath}
\usepackage{amssymb}
\usepackage{graphicx}
\usepackage{wasysym}
\usepackage{natbib}
\bibpunct{(}{)}{;}{a}{}{,} 

\newcommand{\deb}[2]{\displaystyle \frac{\mathrm{d} #1}{\mathrm{d} #2}}

\newcommand{\pdb}[2]{\displaystyle \frac{\partial #1}{\partial #2}}

\newcommand{\dd}[1]{\mathrm{d}^{#1}}
\newcommand{\grad}[1]{\vf{\nabla}#1}
\newcommand{\vf}[1]{{\bf{#1}}}

\newcommand{\Btheta}{B_{\vartheta}}
\newcommand{\Bz}{B_{z}}

\newcommand{\ppsi}{\partial_{\psi}}
\newcommand{\ptheta}{\partial_{\vartheta}}

\newcommand{\kp}{\kappa_{\mathrm{p}}}
\newcommand{\kt}{\kappa_{\mathrm{t}}}

\newcommand{\ez}{\vf{e}_{z}}
\newcommand{\e}{\mathrm{e}}

\newcommand{\mcO}{\mathcal{O}}

\begin{document}

\title{Toward detailed prominence seismology}
\subtitle{I. Computing accurate 2.5D magnetohydrodynamic equilibria}
\author{J.W.S. Blokland\inst{1} \and R. Keppens\inst{2}}
\institute{FOM Institute for Plasma Physics Rijnhuizen, 
           Association EURATOM-FOM, 
	   P.O. Box 1207, 3430 BE Nieuwegein, The Netherlands
           \and
	   Centre for Plasma Astrophysics, Mathematics Department, K.U. Leuven, Celestijnenlaan 200B, 3001 Heverlee, Belgium}
\offprints{J.W.S. Blokland \email{J-W.S.Blokland@Rijnhuizen.nl}}
\date{Received / Accepted}
\abstract{
          Prominence seismology exploits our knowledge of the linear eigenoscillations for representative magnetohydrodynamic models of filaments. To date, 
          highly idealized models for prominences have been used, especially with respect to the overall magnetic configurations.
         }
	 {
          We initiate a more systematic survey of filament wave modes, where we consider full multi-dimensional models with twisted magnetic fields representative 
          of the surrounding magnetic flux rope. This requires the ability to compute accurate 2.5 dimensional magnetohydrodynamic equilibria that balance 
          Lorentz forces, gravity, and pressure gradients, while containing density enhancements (static or in motion).
	 }
	 {
          The governing extended Grad-Shafranov equation is discussed, along with an analytic prediction for circular flux ropes for the Shafranov shift of the 
          central magnetic axis due to gravity. Numerical equilibria are computed with a finite element-based code, demonstrating fourth order accuracy on an 
          explicitly known, non-trivial test case.
	 }
	 {
          The code is then used to construct more realistic prominence equilibria, for all three possible choices of a free flux-function. 
          We quantify the influence of gravity, and generate cool condensations in hot cavities, as well as multi-layered prominences.
	 }
	 {
          The internal flux rope equilibria computed here have the prerequisite numerical accuracy to allow a yet more advanced analysis of the complete spectrum 
          of linear magnetohydrodynamic perturbations, as will be demonstrated in the companion paper.
	 }
\titlerunning{Computing accurate 2.5D MHD equilibria}
\authorrunning{J. W. S. Blokland and R. Keppens}
\maketitle

\section{Prominence seismology and equilibrium configurations \label{sec:intro}}
One of the most fascinating phenomena embedded in the million degree solar coronal plasma is the presence of so-called filaments, which are plasma concentrations 
suspended magnetically against the downward pull of the solar gravitational field. They are up to 100 times colder and denser than their immediate surroundings, 
and in H$\alpha$ observations of the solar disk appear as dark features extending over huge distances. High resolution observations indicate that the filament is 
actually composed of many individual threads, down to the resolution limit of modern observations (about 100 km in width, see~\cite{Linetal05}), while their length 
can reach several tens of megameters. When viewed at the solar limb, the filaments can be identified as prominences, and the surrounding magnetic geometry 
introduces a classification supported by early analytical magnetohydrostatic models~\citep[see e.g.][]{priest88}: normal and inverse polarity prominences differ in 
the relative orientation of the filament-carrying flux rope with respect to the underlying (and overarching arcade) magnetic orientation. Comparing their locations 
to photospheric magnetograms, the filament threads appear primarily suspended above a magnetic neutral line. The study of solar filaments still poses many 
contemporary challenges to theoretical solar physicists, in terms of their sudden formation, their potential to survive for months on end, and their small as well 
as large-scale dynamics. Reviews of the physics of solar prominences include one by~\cite{Labrosse10}, providing an overview of the prominence/filament thermodynamic 
parameters obtained by modern spectroscopic inversions, and another by~\cite{Mackay10}, which compiled insights from essentially magnetohydrodynamic modeling 
that were compared with observations. 

A particularly active research area is known as prominence seismology, whose ultimate goal is the use of observed oscillatory motions in filaments to deduce their 
internal properties, by matching computed with detected frequencies, damping rates, and possibly spatio-temporal eigenstructure of the vibration modes. 
As part of this effort, the review by~\cite{Mackay10} collects insights from observations and theory, and points out how recent works account for the 
multi-threaded fine structure~\citep{Luna10}, as well as for non-adiabatic modifications to the wave modes in prominences~\citep{Soler08}. Another review 
by~\cite{Arregui11} concentrates on how the linear MHD modes can help in explaining the observed attenuation. To make analytic progress, virtually all of 
these efforts make dramatic simplifications about the overall geometry of the magnetic field and thread characteristics: cylindrical flux tubes or 
segments embedded in uniform, thread-aligned magnetic fields form a recurring ingredient. When accounting for internal structure, density variations can explain the 
observed damping of global eigenoscillations in terms of resonant absorption and/or partial ionization, where global kink modes couple to flux-surface localized Alfv\'en 
modes~\citep{Soler10}. Despite the model restrictions, various indicative relations between thread-corona density contrast on the one hand, and mode damping rate to 
period ratio on the other hand, have been established by means of combined analytical and numerical techniques (see e.g.~\citet{Goossens10}). At the same time, it is 
as yet unclear whether these relations generalize to the more realistic topologies where magnetic shear, gravity and pressure variations are all incorporated. 

A major obstacle on the road to generalizing these findings to more realistic prominence models, is that generating accurate equilibrium models for 
prominence-bearing flux ropes becomes a non-linear problem of its own. Even when we restrict the problem to a 2.5D magnetohydrostatic configuration by assuming 
translational invariance along the prominence axis, relatively scarce cases exist where an analytic solution is known. Notable exceptions form the early work 
by~\cite{Dungey53}, where for isothermal conditions the static force balance between the Lorentz force, the (uniform) gravity, and the pressure gradients could be solved 
analytically. \cite{Low_2004} developed under polytropic pressure-density relations, magnetostatic solutions for both normal and inverse polarity 
quiescent prominences. These latter authors pointed out how their construction avoids the free-boundary problem posed by matching the internal flux-rope solution, 
to the external (nearly potential) atmospheric field. In practice, this was achieved by prescribing the prominence to be embedded in a horizontal, circular cylinder, 
whose boundary then coincides with the internal-external transition. In~\cite{Petrie_2007}, this analytic approach was complemented and generalized with a numerical 
treatment where again both polarity types were realized, by iteratively solving the governing elliptic PDE which does fully account for the free-boundary problem. 
It was then demonstrated that a solution can be constructed with the characteristic, observed three-part structure of a cool dense prominence, surrounded by a cavity, 
within a flux rope in a hot coronal environment.

We here revisit the problem of generating accurate, numerical equilibria that are representative of helical flux ropes, supporting dense prominences, possibly showing 
axial (shear) flows. The need to demonstrate high accuracy for the obtained numerical equilibria is a prerequisite for any further analysis quantifying the 
magnetohydrodynamic wave modes. This is common knowledge in fusion-related plasma configurations, where MHD spectroscopy~\citep{Goedbloed_1993} needs to take full 
account of the detailed magnetic geometry to chart out, and eventually invert from, the complete MHD spectrum of waves and instabilities. Meanwhile, this 
approach has already been transported to astrophysically relevant accretion tori~\citep{Goedbloed_2004A,Blokland_2007B}, where a toroidally caged, axisymmetric plasma 
torus has both toroidal (Keplerian-type) and possibly also poloidal flows in a delicate force balance with a central external gravitational field. The MHD spectroscopic 
determination of the eigenoscillations of these accretion tori brought out how flux-surface localized eigenmodes may become unstable as a result of intricate mode 
couplings resulting from the poloidal cross-sectional variation. We here attempt to demonstrate the same ability to chart out all eigenmodes for realistically 
structured prominences, starting with the determination of the continuous parts of the MHD eigenspectrum. These MHD continua form the basic organizing 
structure of the entire collection of MHD eigenmodes~\citep{Book_Goedbloed_2004,Book_Goedbloed_2010}, and because of the multi-dimensional nature of the 
prominence-containing flux rope, will contain avoided crossings and frequency gaps~\citep{Poedts_1991}. To quantify these mode couplings, and to 
subsequently take the next step towards full diagnosis of all global discrete eigenmodes, we must ensure that our equilibrium solution is amenable to such an analysis.

This paper is therefore organized as follows. We begin in Sect.~\ref{sec:equilibrium} with reducing the governing equations for a translationally invariant, 
2.5D magnetohydrostationary configuration to the extended Grad-Shafranov equation. This section recapitulates how 2D equilibria balancing Lorentz force, gravity, and 
pressure gradients allow for axial (shear) flow, and are in principle realizable for three categories of solutions, where either density, entropy, or temperature take 
on a constant value on a single flux surface. We point out the essential role of straight field line coordinates, which must also be constructed from the obtained 
solution. In Sect.~\ref{sec:gravityexpansion}, we obtain the analytical equilibria valid for small gravity, where the outer flux surface is assumed to be circular, and 
the internal, nested flux surfaces are then displaced circles, shifted downwards under the influence of gravity. This is completely analogous to the 
Shafranov shift~\citep{Shafranov_1958} in the axisymmetric tokamak configuration induced by toroidal curvature. This can serve as a clear accuracy test for the 
numerically generated equilibria. In Sect.~\ref{sec:FINESSE}, we briefly explain the MHD equilibrium code FINESSE~\citep{Belien_2002}. This code is used to generate
numerical equilibria. Finally, Sect.~\ref{sec:results} then uses FINESSE and demonstrates how we obtain fourth-order accurate solutions by recovering the analytical case 
presented by~\cite{Dungey53}. It then continues to show the various types of solutions, how they obey the analytically predicted shift for small gravity, and then surveys 
those cases that are no longer tractable by analytic means. This includes a case where at sufficiently large gravity, multiply-layered prominence condensations can exist 
in a helical flux rope. We restrict these computations to cases for given outer cross-sectional shape resulting in only the internal solution. This restriction is needed
because the follow-up analysis of diagnosing the wave modes intrinsically depends on the presence of closed, nested magnetic flux surfaces. The wave mode analysis will
be presented in an accompanying paper~\citep{Blokland_2011B}, where analytic as well as numerical results on the MHD continua for the 2.5D 
equilibria presented here are analyzed.

\section{Solar prominence equilibrium \label{sec:equilibrium}}
We begin our analysis by considering a translational symmetric solar prominence. For this kind of symmetry, Cartesian coordinates $(x,y,z)$ are 
the natural choice. We note that for this symmetry, the equilibrium quantities depend only on the poloidal coordinates~$x$ and $y$. The prominence equilibrium itself 
is modeled by the ideal MHD equations
\begin{align}
  \label{eq:momentum}
  \rho\pdb{\vf{v}}{t} & = -\rho\vf{v}\cdot\grad{\vf{v}} - \grad{p} + \vf{j}\times\vf{B} - \rho\grad{\Phi}, \\
  \label{eq:entropy}
  \pdb{p}{t}          & = -\vf{v}\cdot\grad{p} - \gamma p\grad{}\cdot\vf{v}                       , \\
  \label{eq:induction}
  \pdb{\vf{B}}{t}     & =  \grad{}\times\left(\vf{v}\times\vf{B}\right)                           , \\
  \label{eq:mass}
  \pdb{\rho}{t}       & = -\grad{}\cdot\left(\rho\vf{v}\right)                                    ,
\end{align}
where $\rho$, $p$, $\vf{v}$, $\vf{B}$, $\Phi$, and $\gamma$ are the density, pressure, velocity, magnetic field, gravitational potential, and the ratio of the specific 
heats, respectively. The current density~$\vf{j} = \grad{}\times\vf{B}$ and the equation $\grad{}\cdot\vf{B}=0$ has to be satisfied. We note that the presented ideal
MHD equations are in the non-dimensional form. The equilibrium will be assumed to be time-independent. Furthermore, the thermal pressure and the 
density are related by the ideal gas law $p = \rho T$. From the equations~$\grad{}\cdot\vf{B}=0$ and~$\grad{}\cdot\vf{j}=0$, it follows that the magnetic field and 
the current density can be written as
\begin{align}
  \vf{B} & =  \ez\times\grad{\psi} + \Bz\ez,        \\
  \vf{j} & = -\ez\times\grad{I}    + j_{z}\ez,
\end{align}
respectively, where, $2\pi\psi$ is the poloidal flux and the poloidal stream function~$I = \Bz$. We restrict ourselves to prominence equilibria with a purely axial flow
\begin{equation}
  \vf{v} = v_{z}\ez.
  \label{eq:velocity}
\end{equation}
In this case, Eqs.~\eqref{eq:entropy} and~\eqref{eq:mass} are trivially satisfied. The axial velocity~$v_{z}$ is related to the electric field
\begin{equation}
  \vf{E} = -\vf{v}\times\vf{B} = v_{z}\grad{\psi}.
\end{equation}
From the induction equation presented in Eq.~\eqref{eq:induction}, it follows that $v_{z} = v_{z}(\psi)$. Since axial flows are routinely observed in prominence 
threads, we include the parameter $v_{z}$ in our analysis. Owing to the translational symmetry, it does not play any role in the actual force balance, but we show 
in an accompanying paper~\citep{Blokland_2011B} how the velocity shear in the equilibrium modifies the eigenspectrum.

The momentum equation Eq.~\eqref{eq:momentum} can be projected in three ways. The first one is in the axial direction, which shows that the poloidal stream function
is a flux function, i.e. $I=I(\psi)$. The second projection is parallel to the poloidal magnetic field resulting in two equations
\begin{equation}
  \begin{aligned}
    \left. \pdb{p}{x} \right|_{\psi=\mathrm{const}} & = -\rho \pdb{\Phi}{x},      \\
    \left. \pdb{p}{y} \right|_{\psi=\mathrm{const}} & = -\rho \pdb{\Phi}{y}.
  \end{aligned}
  \label{eq:parallelBpol}
\end{equation}
These two equations have to be satisfied simultaneously. We conclude from these that the pressure $p = p(\psi; x,y)$. The third and last projection is
the one perpendicular to the poloidal magnetic field. This leads to the extended Grad-Shafranov equation
\begin{equation}
  \grad{}^{2}\psi = - I\deb{I}{\psi} - \pdb{p}{\psi}.
  \label{eq:gradshafranov}
\end{equation}

\subsection{Further reduction for specific flux functions}
The two equations parallel to the poloidal magnetic field given in Eq.~\eqref{eq:parallelBpol} can be solved analytically under the extra assumption that either the 
temperature~$T$, the density~$\rho$, or the entropy~$S=p\rho^{-\gamma}$ is a flux function. The assumption that the temperature is a flux function can be justified 
because of the high thermal conductivity along the magnetic field lines. The resulting pressure can then be written as
\begin{equation}
  p(\psi;x,y) = p_{0}(\psi) \exp \left[ -\frac{\Phi(x,y)}{T(\psi)} \right],
  \label{eq:pressureT}
\end{equation}
where $p_{0}$ is the pressure for a static pure Grad--Shafranov equilibrium without gravity. The extended Grad-Shafranov equation in Eq.~\eqref{eq:gradshafranov} reduces 
to
\begin{equation}
    \grad{}^{2} \psi = 
      -I\deb{I}{\psi} - \left[ \deb{p_{0}}{\psi} + \frac{p_{0} \Phi}{T^{2}}\deb{T}{\psi} \right] \, \exp \! \left[ - \frac{\Phi}{T} \right].
  \label{eq:gradshafranovT}
\end{equation}
This extended Grad--Shafranov equation is a minor generalization of the one presented in \citet{Petrie_2007}, by allowing a fully general external gravitational potential. 
This case where temperature is a flux function is most relevant for the quiescent solar prominences, which are long-lived structures and hence will be temperature 
equilibrated on flux surfaces. The isothermal case forms a subset of this solution class.

Another possibility, on MHD timescales, is to assume that the density is a flux function. In this case, the pressure reads
\begin{equation}
  p(\psi;x,y) = p_{0}(\psi) \left[ 1 - \frac{\Phi(x,y)}{T_{\rho}(\psi)} \right],
  \label{eq:pressurerho}
\end{equation}
where the quasi-temperature~$T_{\rho} \equiv p_{0} / \rho$. Under this assumption, the extended Grad-Shafranov equation in Eq.~\eqref{eq:gradshafranov} can be written as
\begin{equation}
  \begin{aligned}
    \grad{}^{2}\psi & = -I\deb{I}{\psi} \\
                    & \quad 
                        - \left[ \deb{p_{0}}{\psi} + \frac{p_{0} \Phi}{T_{\rho}^{2}}\deb{T_{\rho}}{\psi} \left( 1 - \frac{\Phi}{T_{\rho}} \right)^{-1} \right]
                        \left[ 1 - \frac{\Phi}{T_{\rho}} \right].
  \end{aligned}
  \label{eq:gradshafranovrho}
\end{equation}
This case where density is a flux function will be shown in our companion paper to be susceptible to unstable continuum modes, under quantifiable stability 
criteria~\citep{Blokland_2011B}. Hence, it might be relevant to the short-lived active region prominences, where these instabilities may relate to the sudden 
disappearance of filaments.

The final option is to assume that the entropy~$S$ is a flux function, as also discussed in \citet{Petrie_2007}.
We again generalize it slightly by adopting an arbitrary potential, for which the pressure reads
\begin{equation}
  p(\psi;x,y) = p_{0}(\psi) \left[ 1 - \frac{\gamma-1}{\gamma}\frac{\Phi(x,y)}{T_{S}(\psi)} \right]^{\gamma / (\gamma-1)},
  \label{eq:pressureS}
\end{equation}
and the extended Grad-Shafranov equation in Eq.~\eqref{eq:gradshafranov} reduces to
\begin{equation}
  \begin{aligned}
    \grad{}^{2}\psi & = -I\deb{I}{\psi} \\
                    & \quad
                        - \left[ \deb{p_{0}}{\psi} + \frac{p_{0} \Phi}{T_{\rho}^{2}}\deb{T_{S}}{\psi} \left( 1 - \frac{\gamma-1}{\gamma}\frac{\Phi}{T_{S}} \right)^{-1} \right] \\
                    & \quad \phantom{-} \quad
                        \left[ 1 - \frac{\gamma-1}{\gamma}\frac{\Phi}{T_{S}} \right]^{\gamma/(\gamma-1)} \,,
  \end{aligned}
  \label{eq:gradshafranovS}
\end{equation}
where the quasi-temperature~$T_{S} \equiv S\rho_{0}^{\gamma-1}$. We note that all three cases can also be derived using the expressions for axisymmetric accretion tori 
with a purely toroidal flow, as presented in the paper by \citet{Blokland_2007B}, by setting $R=R_{0}=1$ in their formulae. This case where entropy is a flux function 
is the natural one to consider when also plasma rotation would be incorporated. 

The equation for the density for all three cases can easily be derived by inserting the corresponding pressure equation into the equation for the momentum parallel to 
the poloidal magnetic field lines given in Eq.~\eqref{eq:parallelBpol}. The resulting equation is
\begin{equation}
  \rho(\psi;x,y) = \rho_{0}(\psi) \times
  \begin{cases}
    \exp \! \left[ - \dfrac{\Phi}{T}  \right]                                          & \text{for }T=T(\psi)        \\
    1                                                                                  & \text{for }\rho=\rho(\psi)  \\
    \left[ 1 - \dfrac{\gamma -1}{\gamma} \dfrac{\Phi}{T_{S}} \right]^{1/(\gamma - 1)}  & \text{for }S=S(\psi)
  \end{cases}.
\end{equation}
Here, the flux function~$\rho_{0}$ corresponds to the density of a related static equilibrium without gravity.

\subsection{Straight field line coordinates}
As we will adopt them for the actual stability analysis in our accompanying paper, we briefly discuss the `straight field line' coordinates.
These coordinates are an essential ingredient of an accurate stability analysis. For the conversion from the
Cartesian $(x,y,z)$ to straight field line coordinates $(x^{1}\equiv\psi, x^{2}\equiv\vartheta, x^{3}\equiv z)$, one needs the metric 
tensor and the Jacobian associated with the non--orthogonal coordinates in which the equilibrium field lines appear to be straight. 
Such a transformation is standard practice in MHD stability studies for laboratory tokamak plasmas. The metric elements~$g_{ij}$ and 
the Jacobian~$J$ are
\begin{equation}
  \begin{aligned}
    g^{ij} & = \grad{x^{i}}                          \cdot \grad{x^{j}},                           & \quad
    g_{ij} & = \frac{\partial\vf{r}}{\partial x^{i}} \cdot \frac{\partial\vf{r}}{\partial x^{j}},  \\
    J      & = \left( \grad{\psi}\times\grad{\vartheta} \cdot \grad{z} \right)^{-1},               & \quad & &
  \end{aligned}
\end{equation}
respectively. Here, the poloidal angle~$\vartheta$ is constructed such that the magnetic field lines are straight in the $(\vartheta, z)$--plane. 
The slope of these lines is a flux function
\begin{equation}
  \left. \deb{z}{\vartheta} \right|_{\mathrm{field line}} = 
  \frac{\vf{B}\cdot\grad{z}}{\vf{B}\cdot\grad{\vartheta}} = JI = q(\psi),
  \label{eq:safetyfactor}
\end{equation}
where $q$ is the safety factor. Comparing this expression with the one for tokamak plasmas \citep{Wesson_2004}, 
one should realize that for tokamak plasmas the safety factor $q$ is dimensionless, while here the factor $q$ has a length dimension.
As for tokamak plasmas \citep{Goedbloed_1975,vanderHolst_2000B,Blokland_2007B}, we introduce an expression for the poloidal curvature of the magnetic surfaces
\begin{alignat}{2}
  \kp & = -\vf{n} \cdot \left( \vf{t}\cdot\grad{\vf{t}} \right)                     &  
      & = \frac{1}{J}\left( \ppsi - \ptheta\frac{g_{12}}{g_{22}} \right) J\Btheta \,,  
\end{alignat}
where the unit vectors $\vf{n} = \grad{\psi} / |\grad{\psi}|$ and $\vf{t} = \vf{B}_{\vartheta} / \Btheta$, and~$\Btheta$
is the poloidal magnetic field. The toroidal curvature $\kt$ that is present in actual tokamak equilibria, is of course zero for a translational symmetric equilibrium.
It is important to realize that the straight field coordinates can only be constructed when the solution~$\psi(x,y)$ has been computed from the extended 
Grad-Shafranov equation given in Eq.~\eqref{eq:gradshafranov}.

\section{Small gravity expansion \label{sec:gravityexpansion}}
In the previous section, we derived equations for the prominence equilibrium. In this section, we quantify the effect of gravity by means of a small gravity 
expansion. We will demonstrate in our companion paper that gaps will appear in the continuous MHD spectrum because of mode coupling, which is the 
result of the presence of gravity. This kind of expansion is similar to the small inverse aspect ratio $\epsilon = a/R_{0}$ expansion for tokamak plasmas, where $a$ 
and $R_{0}$ are the minor radius of the plasma and the geometry axis of the tokamak, respectively \citep{Shafranov_1958}. Mathematically,
both expansions, small gravity expansion and the small inverse aspect ratio expansion, are Taylor expansions. In the remaining part of this paper, we assume the 
following gravitational potential in which the prominence is embedded
\begin{equation}
  \Phi(x,y) = (x-x_{0})g,
  \label{eq:gravity}
\end{equation}
where $x_{0}$ is the location of the center of the last closed flux surface of the prominence and the gravity is represented by the constant $g$.

The solar prominence equilibrium is expanded assuming that the gravity is small and that the outer flux surface is circular. Using these approximations,
the flux surfaces can be represented by slightly displaced circles (\cite{Shafranov_1958}), which allows for the exploitation of non-orthogonal polar 
coordinates $(r,\theta,z)$, where $r$ and $\theta$ are the radius and the polar angle, respectively. Up to first order, we 
approximate
\begin{equation}
  \begin{aligned}
    \psi(x,y) & = \psi(r),                            \\
    x         & = x_{0} + r\cos(\theta) - \Delta(r),  \\
    y         & = r\sin(\theta),
  \end{aligned}
  \label{eq:coordinates}
\end{equation}
where $\Delta(r)$ is the Shafranov shift (\cite{Shafranov_1958}), which is expected to be in the downwards direction and caused by the gravity.  
As mentioned before, these polar coordinates are non-orthogonal and the associated metric elements are
\begin{equation}
  \begin{aligned}
    h_{11} & \approx 1 - 2\Delta' \cos(\theta),  \quad &   h_{22} & \approx r^{2},                      \\
    h_{12} & \approx r\Delta' \sin(\theta),            &   h_{33} & =       1,
  \end{aligned}
\end{equation}
which means that the Jacobian $J \approx r [ 1 - \Delta' \cos(\theta) ]$.

Using the polar coordinates and expanding the extended Grad-Shafranov equation given in Eqs.~\eqref{eq:gradshafranovT},~\eqref{eq:gradshafranovrho}, 
and~\eqref{eq:gradshafranovS} up to first order leads to one equation for the equilibrium ($\mcO(1)$) and one equation for the Shafranov shift ($\mcO(g)$)
for all three cases. The $\mcO(1)$ equilibrium relation is
\begin{equation}
  \deb{}{r} \!\! \left[ p_{0} + \tfrac{1}{2}\left( B_{\theta}^{2} + \Bz^{2} \right) \right] + \frac{B_{\theta}^{2}}{r} = 0,
  \label{eq:equilibrium_cylindrical}
\end{equation}
where $B_{\theta} \equiv \psi'$ is the poloidal magnetic field expressed in polar coordinates, 
where the prime indicates the derivative with respect to the radius $r$. This is exactly as expected, since this equation merely expresses the force balance in a cylinder. 
From this equation, one can see that the pressure and the magnetic field components are of the same order $\mcO (1)$. This means that
the plasma beta $\beta = 2p / B^{2} = \mcO(1)$. This is very different from tokamak physics where the pressure is two orders of magnitude lower than 
the toroidal magnetic field component, which ensures that the plasma beta there is of the order $\mcO(\epsilon^{2})$ (\cite{vanderHolst_2000B}).
For such tokamak plasmas, the Shafranov shift is outwards due to the toroidicity, pressure, and toroidal flow (\citet{vanderHolst_2000B}).

The equation for the Shafranov shift in prominence equilibria is
\begin{equation}
  \deb{\Delta}{r} = \frac{1}{r B_{\theta}^{2}} \int_{0}^{r} r^{2}g \deb{\rho_{0}}{r} \dd{}r,
  \label{eq:shafranovshift}
\end{equation}
which shows that the magnetic axis is shifted downwards because of the gravity. We note that for zero gravity or a constant density there will be no shift 
at all, again in accord with the cylindrical configuration then expected. 

As mentioned before, to perform the spectral formulation and compute the continuous MHD spectrum, the analysis is done in straight field line coordinates. Exploiting the
small gravity expansion, the straight field line metric elements can be approximated by
\begin{equation}
  \begin{aligned}
    g_{11} & \approx \frac{1}{{\psi'}^{2}}\left[ 1 - 2\Delta' \cos (\vartheta) \right],  \quad &
      g_{22} & \approx r^{2} \left[ 1 + 2\Delta' \cos (\vartheta) \right],                     \\
    g_{12} & \approx \frac{r}{\psi'} \left( r\Delta' \right)' \sin (\vartheta),                &
      g_{33} & = 1.
  \end{aligned}
  \label{eq:smallg_metric_g}
\end{equation}
In deriving these expressions, we exploited that the safety factor $q$ from Eq.~\eqref{eq:safetyfactor} is a flux function and therefore
the Jacobian $J (= r / \psi')$ has to be a flux function. This allows us to find a relationship between the polar angle $\theta$ and the straight 
field line angle $\vartheta$
\begin{equation}
  \theta \approx \vartheta + \Delta' \sin(\vartheta).
  \label{eq:smallg_theta}
\end{equation}

\section{The equilibrium code FINESSE \label{sec:FINESSE}}
\subsection{Code basics: discretization}
The FINESSE code was originally developed by \citet{Belien_2002} and designed to solve the coupled generalized Grad-Shafranov equation together with the
algebraic Bernoulli equation (\cite{Hameiri_1983,Zelazny_1993,Goedbloed_2004A}). The design of the code is such that we could easily extend it by implementing the extended 
Grad--Shafranov equations given in Eqs.~\eqref{eq:gradshafranovT},~\eqref{eq:gradshafranovrho}, and~\eqref{eq:gradshafranovS}, as realized in version 1.3 of the FINESSE code. 
These equations are solved for a given poloidal cross-section using the Picard iteration scheme in combination with a standard matrix solver. 
The boundary conditions are such that a fixed given boundary shape represents the last closed flux surface. As a discretization scheme, we exploited a finite element method in 
combination with the standard Galerkin method. As elements, we used isoparametric Hermite elements. These elements ensure that the computed solution has the required 
accuracy needed for the stability analysis. 

\subsection{Scaling for prominence equilibria}
The exploited scaling in FINESSE is such that all numerical quantities actually computed are quantified in 1) the profile variations of order unity, and 2) the 
amplitudes that determine the relative strengths. The idea is that solving the governing elliptic PDE numerically should exploit quantities of order unity as much as 
possible. This means for example that we ensure a scaling such that, e.g. the flux function $\psi$ varies from 0 to 1. In the results section (Sect.~\ref{sec:results}), 
we present either dimensionless quantities such as the local plasma $\beta$, or quantify the pressure or the gravity parameter $g$ in code units. Since we vary $g$ in 
what follows, this may seem unusual at first for solar filaments, because one can obviously not alter the solar gravitational field. We therefore here explain the 
inherent scaling.

For this purpose, some typical values for solar prominence conditions taken from~\citet{Labrosse10} and~\citet{Mackay10} are as follows. Lengths are expressed in units of 
the prominence-carrying magnetic loop radius. This length unit $L$ can be varied from 100 km (if a single prominence thread is being modeled) to 10 Mm, the latter being an 
upper width quoted for solar prominence structures. Taking a typical temperature of 8000 K (a factor 100 cooler than the corona) and an electron number 
density $n_e\sim 10^{16}\,{\mathrm m}^{-3}$, the density is then $\rho\sim 1.673  \times 10^{-11}\, {\mathrm {kg}} \,{\mathrm m}^{-3}$ and a reference pressure value 
is $p\sim 0.00221 \, {\mathrm N}\,{\mathrm m}^{-2}$. We note that the sound speed can then be estimated to be $15\, {\mathrm {km}}\,{\mathrm s}^{-1}$. If we also use a typical 
value of magnetic field strength $B\sim 10^{-3} \, {\mathrm{T}}$, the Alfv\'en speed is $218 \, {\mathrm{km}}\,{\mathrm{s}}^{-1}$. For these parameters, the plasma beta 
can be as low as $\beta=0.0055$, but we note that values of up to order 0.1 can be deduced from observations. The gravitational acceleration at a solar radius is of 
order $g\sim 272 \, {\mathrm N}\,{\mathrm {kg}}^{-1}$, and a typical hydrostatic scale-height inferred from these values combined, yields 
$\Lambda=p/\rho g\sim 485 \,{\mathrm {km}}$. When we quantify the parameter $g$ in dimensionless units below, we actually mean the value of the dimensionless combination 
(with $\mu_0$ the permeability)
\begin{equation}
 \frac{L \rho \mu_0 g}{B^2} \simeq 0.0057 \,.
\end{equation}
We note that this dimensionless quantification of the strength of the gravitational field can meaningfully be altered by orders of magnitude, as it is directly proportional to 
the size of, and the density in the prominence, and inversely proportional to the square of the field strength. Hence, larger values of the combination represent larger, 
denser prominences in weaker embedding magnetic fields, while the value given above is a reasonable reference value for a typical flux rope. Code units in essence use a 
dimensionalization based on $L$, $\rho$, and $B$, so that for instance the pressure is in units of $B^2/\mu_0\sim 0.795 \, {\mathrm N}\,{\mathrm m}^{-2}$. This means that
the reference pressure value quoted above is reached at $p=0.0028$ in units of $B^{2}/\mu_{0}$.

\section{Numerical results \label{sec:results}}
We now demonstrate that FINESSE can accurately compute prominence equilibria by comparing the numerical solution with the analytical
solution derived by \citet{Dungey53}. We then discuss two classes of equilibria. The first class represents cool prominences embedded in a hot medium
and the second one in combination with proper chosen parameters represents multi-layered prominences.

\subsection{Accuracy test: the Dungey solution}
\begin{figure*}[ht]
  \centering
  \includegraphics[width=0.7\textwidth]{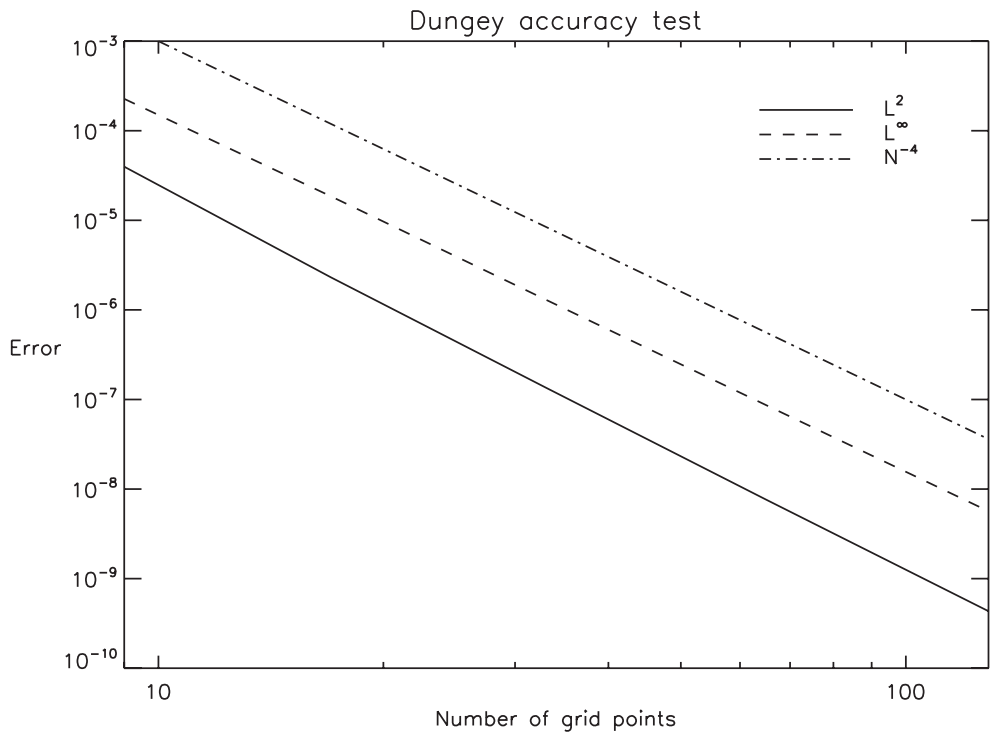}
  \caption{The accuracy test of FINESSE with respect to the analytical solution derived by \citet{Dungey53}. The quantity $L^{2}$ and $L^{\infty}$ are the
           average and maximum error between the Dungey solution and the numerical solution over the whole interior.}
  \label{fig:dungey}
\end{figure*}

To demonstrate FINESSE accurately computes the prominence equilibria, we compare the numerical result with the analytical solution derived by \citet{Dungey53}.
The derived solution
\begin{equation}
  \psi(x,y) = \psi_{1} \left[ \e^{-\alpha x} - 2\e^{-\tfrac{1}{2}\alpha x}\cos \left(\tfrac{1}{2}\alpha y\right) + 1 \right],
\end{equation}
where $\alpha \equiv g / T$, satisfies the extended Grad-Shafranov equation given in Eq.~\eqref{eq:gradshafranovT} with the extra assumption that the temperature $T$ 
is constant. The solution itself is used to specify the last closed surface so FINESSE can compute the interior for various resolutions. The error between the numerical 
solution and the analytical one as a function of the number of grid points is shown in Fig.~\ref{fig:dungey}. The quantities $L^{2}$ and $L^{\infty}$ are the average error 
and the maximum error evaluated over the whole interior. The plot clearly demonstrates that FINESSE shows a fourth order convergence as it was designed to do. Furthermore, 
it also shows that moderate resolution is enough to obtain a highly accurate equilibrium solution.

\subsection{Cool prominence surrounded by a hot medium \label{sec:cool}}
The first equilibrium class of solar prominences is the one where a prominence is embedded in a hot medium. These kinds of equilibria are specified by the 
following profiles: 
\begin{align}
  \label{eq:cool_I}
  I^{2}(\psi) & = A(1 - \tfrac{3}{4} \psi),                     \\
  \label{eq:cool_p}
  p_{0}(\psi) & = AA_{2}(1 - \tfrac{9}{10}\psi),                \\
  \label{eq:cool_T}
  T_{0}(\psi) & = A_{3}(\tfrac{1}{5} + 9\psi^{2} - 6\psi^{3}),  \\
  \label{eq:cool_vz}
  v_{z}(\psi) & = 0,
\end{align}
where $A_{2}=0.1$ and $A_{3}=1$. The amplitude A is the overall amplitude computed by FINESSE as part of the equilibrium solution. The function $T_{0}$ 
represents the (quasi) temperature of the three choices of the flux function, the density, the entropy, or the temperature. The gravity $g$, scaled as explained earlier, 
has been varied from $0.001$ up to $1.000$. For most equilibria of this subsection, the temperature is assumed to be a flux function, except when stated differently. 
Furthermore, we assume a circular boundary. The resolution adopted for the numerical computations are 101 radial points and 129 points in the poloidal direction.
The specified profiles for $I^{2}(\psi)$ and $p_{0}(\psi)$ are of the same form used by \cite{Low_1995}, \cite{Low_2004}, and \cite{Petrie_2007}. The form of 
the~$T_{0}(\psi)$ profile is the same as the one used by \citet{Petrie_2007}, which represents a low temperature in the center while the temperature is high at the edge 
of the prominence.

\begin{figure*}[ht]
  \centering
  \includegraphics[width=0.6\textwidth,trim=3cm 0cm 3cm 0cm]{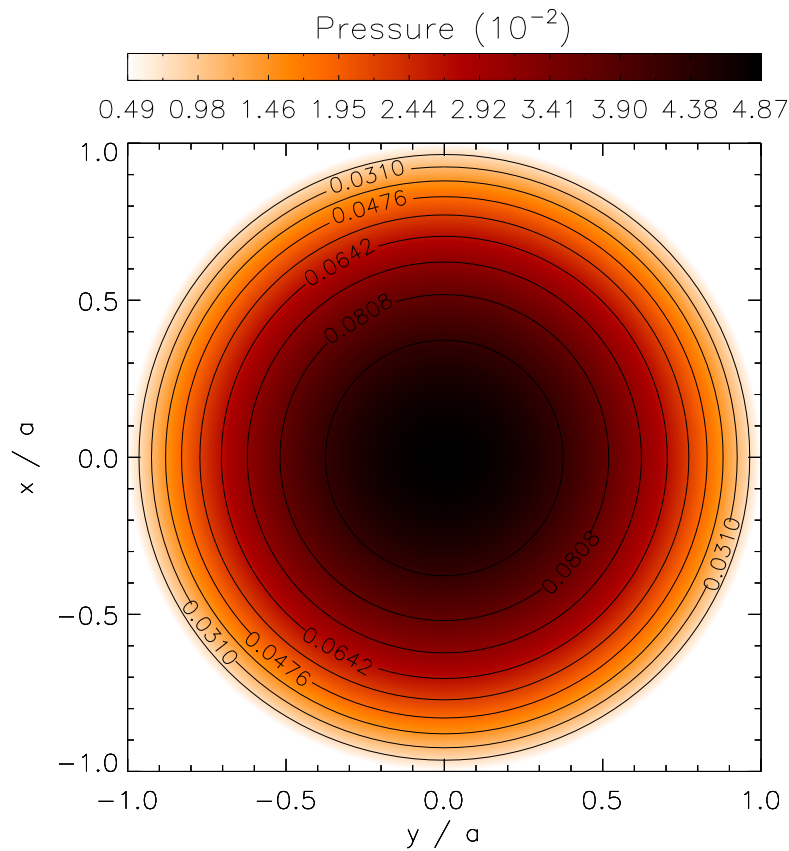}
  \caption{The two-dimensional pressure (red-scale) and plasma beta $\beta=2p/B^{2}$ (contours) profile for a cool solar prominence surrounded by a hot medium with
           a gravity $g=0.001$. The solar surface is below the figure.}
  \label{fig:cool_smallg_pressure}
\end{figure*}

\begin{figure*}[ht]
  \centering
  \begin{tabular}{ll}
    \includegraphics[width=0.5\textwidth]{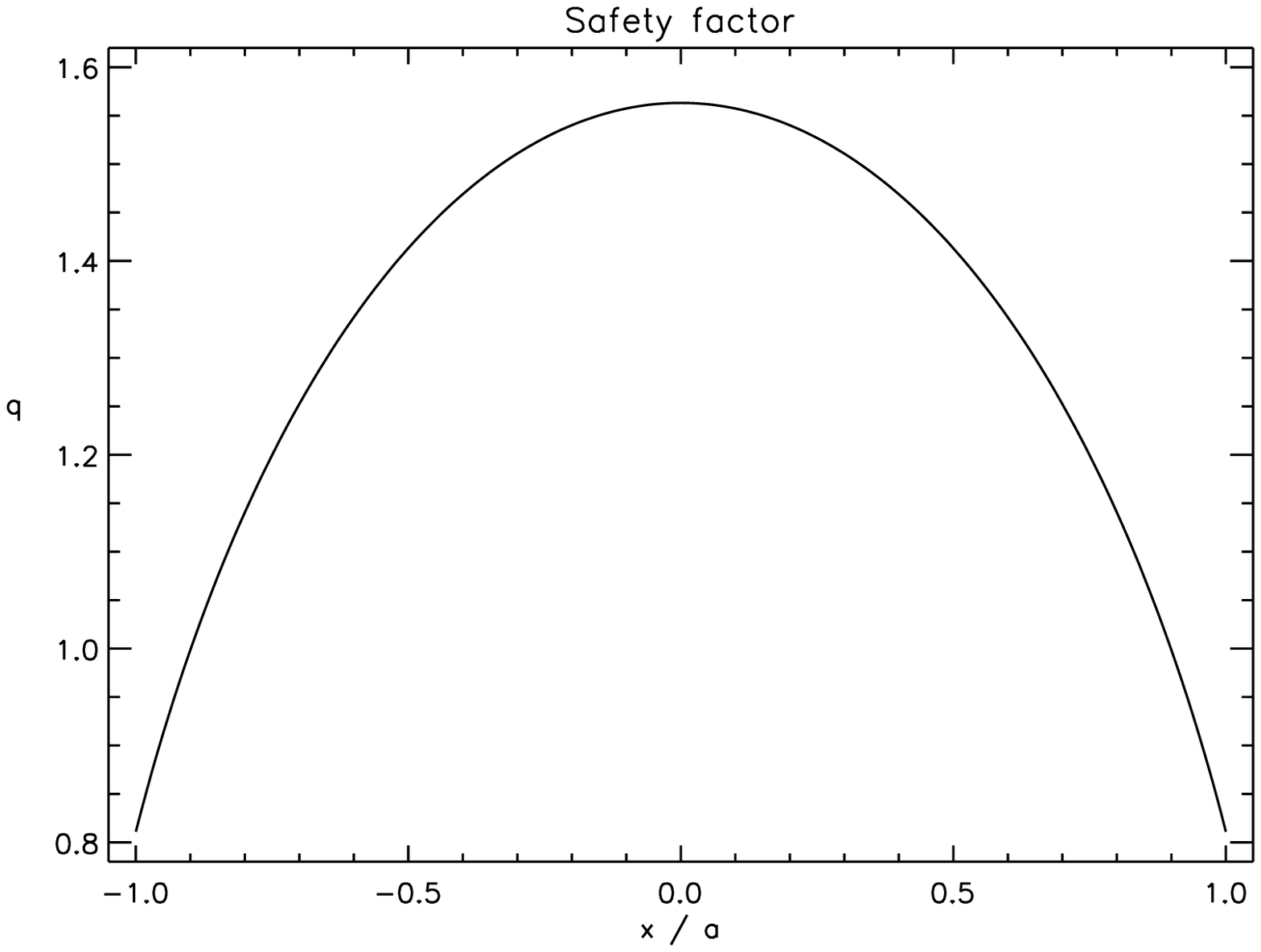}  &
    \includegraphics[width=0.5\textwidth]{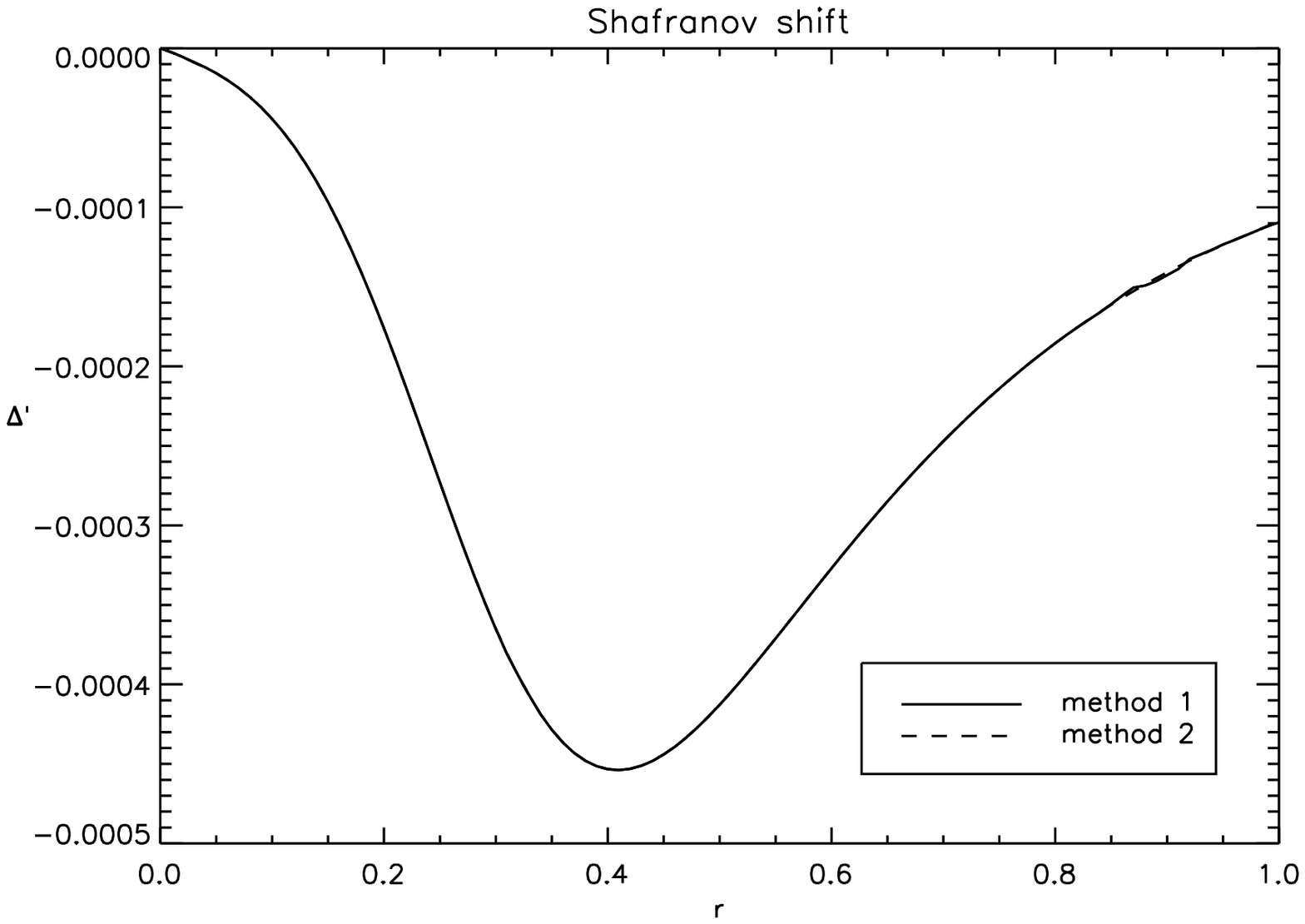}
  \end{tabular}  
  \caption{Left: the safety factor $q$ at the mid-plane for a gravity $g=0.001$.
           Right: the radial derivative of the Shafranov shift $\Delta(r)$ as a function of the radius $r$.}
  \label{fig:cool_smallg_safetyfactor}
\end{figure*}

For the first equilibrium, we set the gravity to be $g=0.001$. The resulting pressure and plasma beta $\beta = 2p/B^{2}$ are shown in Fig.~\ref{fig:cool_smallg_pressure}.
The plot shows clearly a high pressure in the center and a low pressure at the edge of the prominence. A similar observation can be made about the plasma beta, which
is around 0.097 in the core and at the edge has the value of 0.014. The variation in the density, not shown, ranges from 0.006 at the edge up to 1.000 in the centre. 
The central temperature has a value of 0.048, while at the edge it is 0.780, which clearly resembles a cool prominence embedded in a hot medium. For the previously 
mentioned beta values, the magnetic field as well as the thermal pressure will play an important role in the stability analysis of our accompanying paper. In this 
stability analysis the safety factor $q$, shown in Fig.~\ref{fig:cool_smallg_safetyfactor}, is also an essential ingredient. From fusion research, we know that around 
the $q=1$ and $q=3/2$ surface, mode coupling may occur which in its turn can create gaps in the MHD continuous spectrum or even drive the continuous spectrum unstable. 
The plot of the safety factor shows that both surfaces exist in this cool prominence equilibrium. The same figure also shows the radial derivative of the Shafranov shift. 
This derivative was calculated in two different ways. The first applied the second of the equations given in Eq.~\eqref{eq:coordinates} and the ability of
FINESSE to perform the transformation to straight field line coordinates. The second method numerically solved the Shafranov shift equation in 
Eq.~\eqref{eq:shafranovshift} using Simpson's rule (\cite{Press_1988}). The first method is valid for any value of the gravity parameter as long as the flux surfaces 
are circles, while the second method can only be used for small gravity in combination with circular flux surfaces. The RHS plot of 
Fig.~\ref{fig:cool_smallg_safetyfactor} shows excellent agreement between the two described methods.

\begin{figure*}[ht]
  \centering
  \includegraphics[width=0.6\textwidth,trim=3cm 0cm 3cm 0cm]{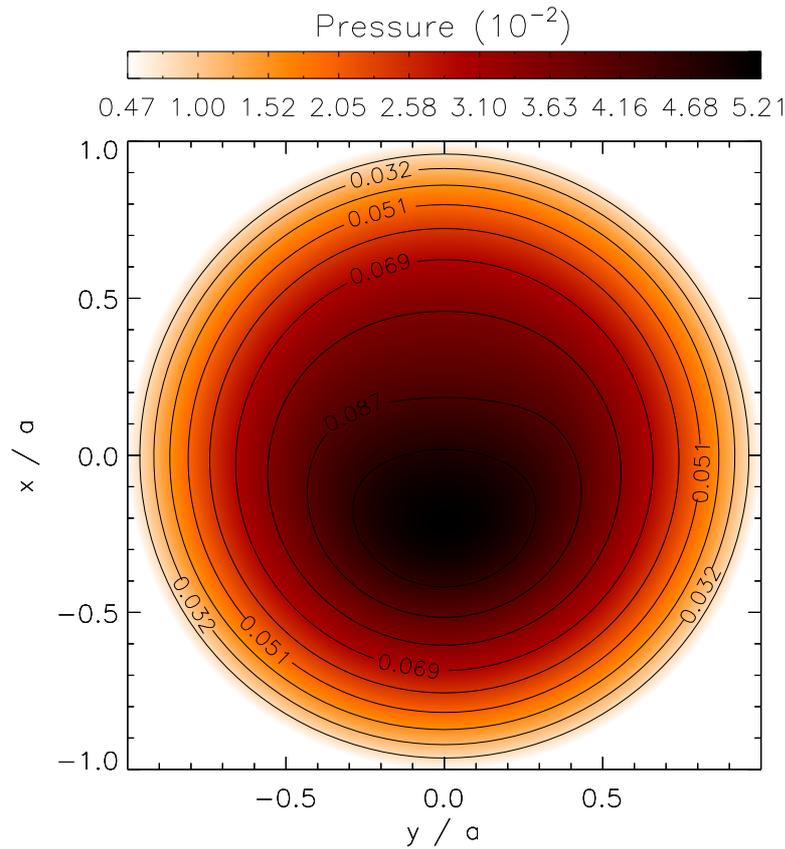}
  \caption{The two-dimensional pressure (red-scale) and plasma beta $\beta=2p/B^{2}$ (contours) profile for a cool solar prominence surrounded by a hot medium with
           a gravity $g=0.100$. The solar surface is below the figure.}
  \label{fig:cool_intermediateg_pressure}
\end{figure*}

\begin{figure*}[ht]
  \centering
  \begin{tabular}{ll}
    \includegraphics[width=0.5\textwidth]{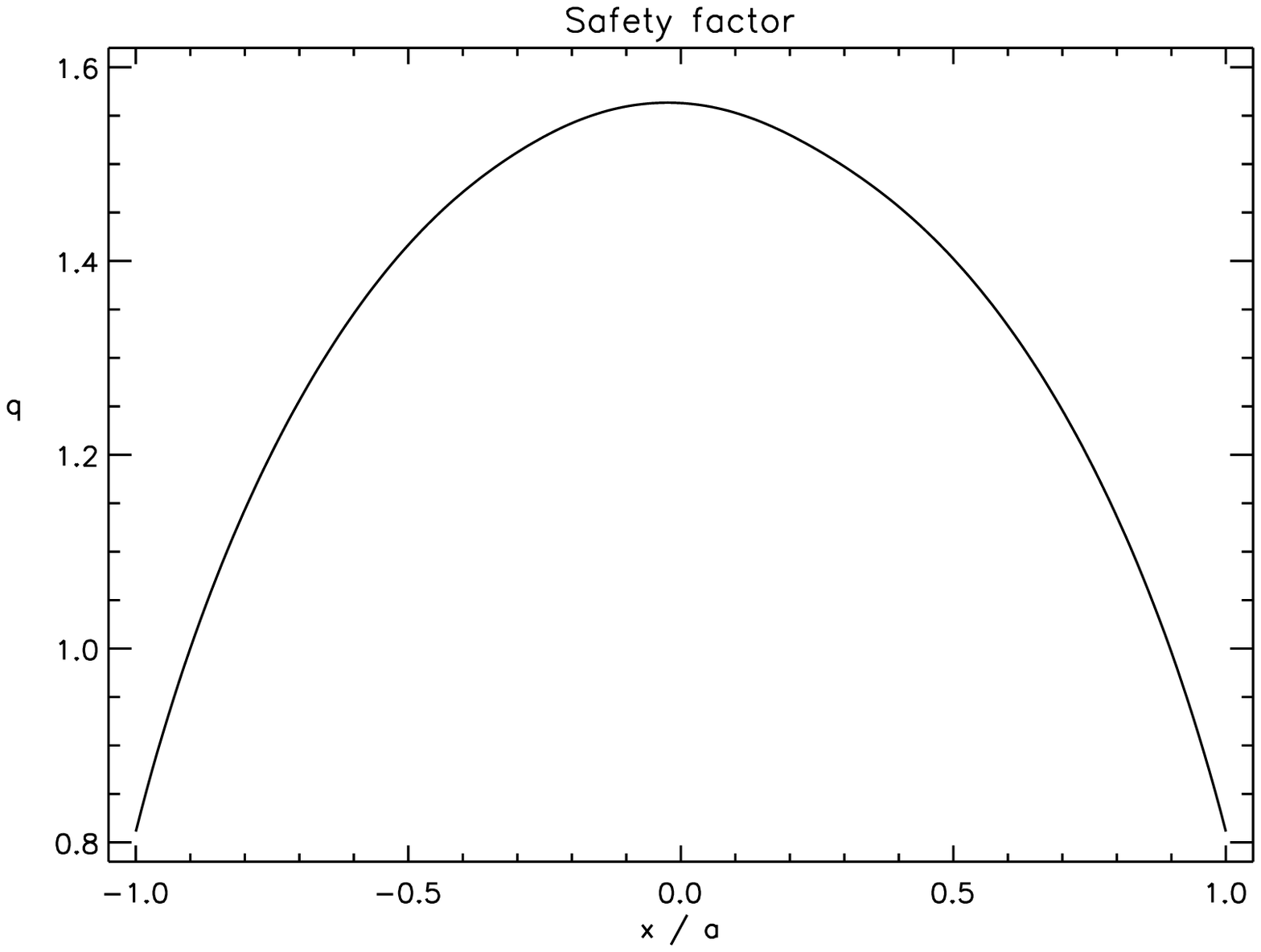}  &
    \includegraphics[width=0.5\textwidth]{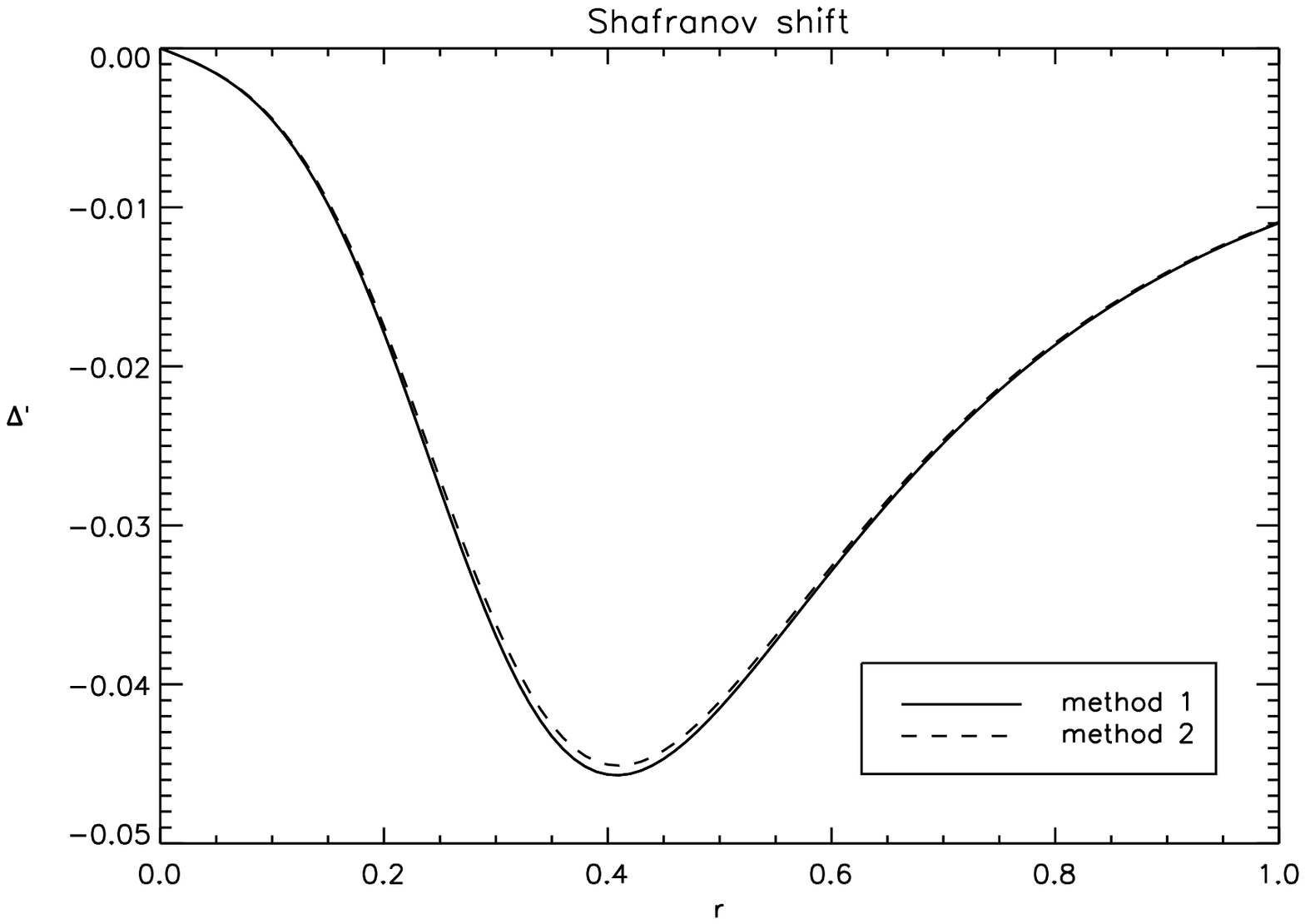}
  \end{tabular}  
  \caption{Left: the safety factor $q$ at the mid-plane for a gravity $g=0.100$.
           Right: the radial derivative of the Shafranov shift $\Delta(r)$ as a function of the radius $r$.}
  \label{fig:cool_intermediateg_safetyfactor}
\end{figure*}

We then increase the gravity parameter by a factor of 100, i.e. $g=0.100$. The pressure and plasma beta are shown in Fig.~\ref{fig:cool_intermediateg_pressure}. Since 
the fact that the parameter $g$ is 100 times stronger, the pressure maximum is shifted downwards as expected. The plasma beta has values similar to those in the previous 
case, its maximum and minimum values being 0.105 and 0.014, respectively. The two-dimensional density profile, not shown here, has a similar behavior to that of the 
pressure profile. Its maximum is also shifted downwards as expected. Fig.~\ref{fig:cool_intermediateg_safetyfactor} shows the safety factor $q$ and the radial derivative 
of the Shafranov shift $\Delta(r)$. Comparing the safety factor with the one of the previous equilibrium, one notices that there is hardly any difference. The two methods 
for computing the Shafranov shift now show a small difference. The second method, which using the Shafranov shift equation in Eq.~\eqref{eq:shafranovshift},
underestimates the radial derivative for all radii.

\begin{figure*}[ht]
  \centering
  \includegraphics[width=0.6\textwidth,trim=3cm 0cm 3cm 0cm]{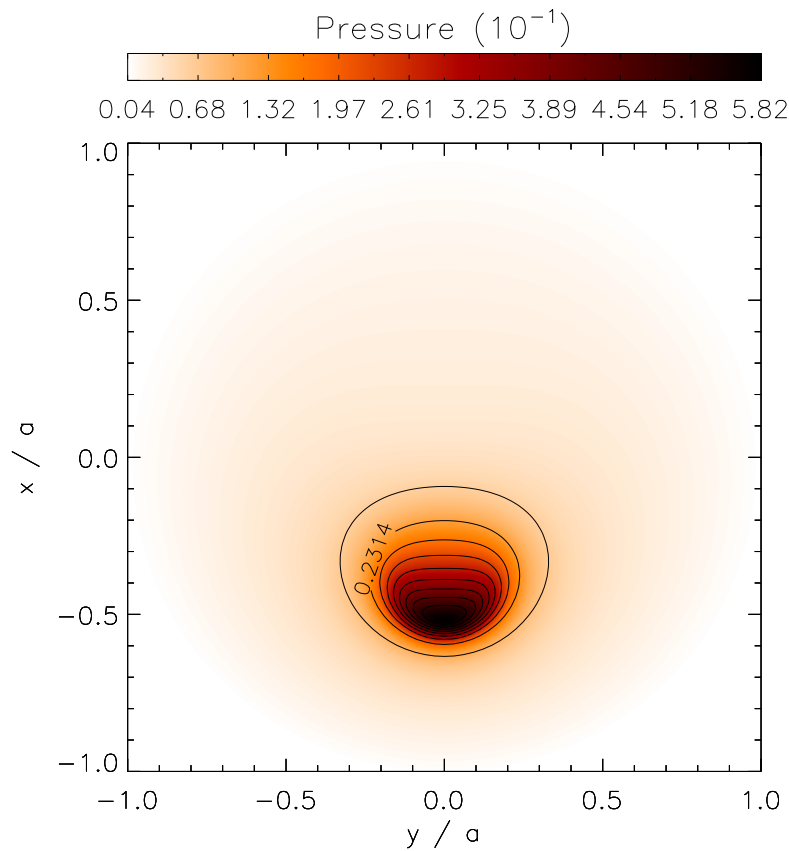}
  \caption{The two-dimensional pressure (red-scale) and plasma beta $\beta=2p/B^{2}$ (contours) profile for a cool solar prominence surrounded by a hot medium with
           a gravity $g=1.000$. The solar surface is below the figure.}
  \label{fig:cool_strongg_pressure}
\end{figure*}

\begin{figure*}[ht]
  \centering
  \begin{tabular}{ll}
    \includegraphics[width=0.5\textwidth]{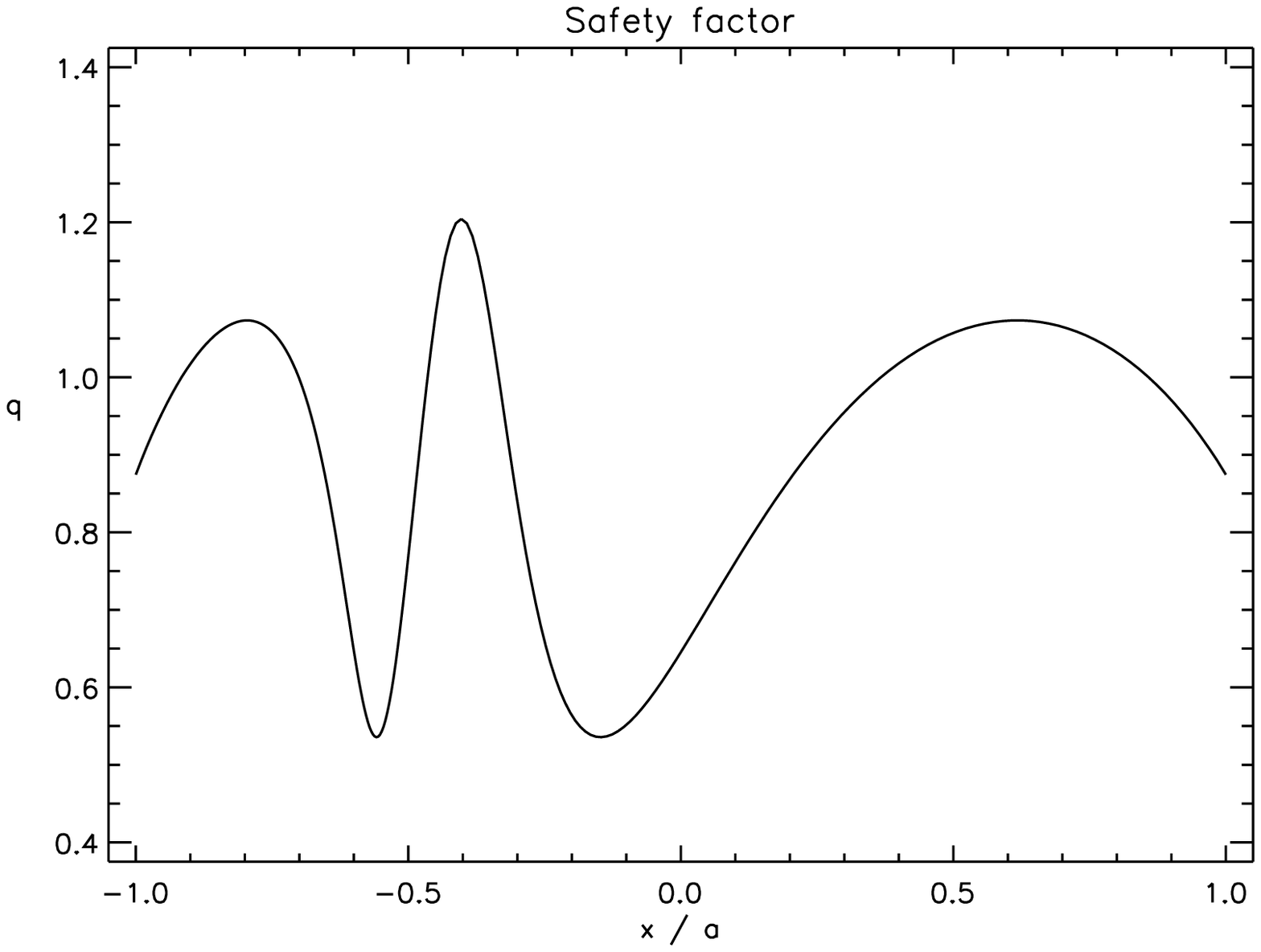} &
    \includegraphics[width=0.5\textwidth]{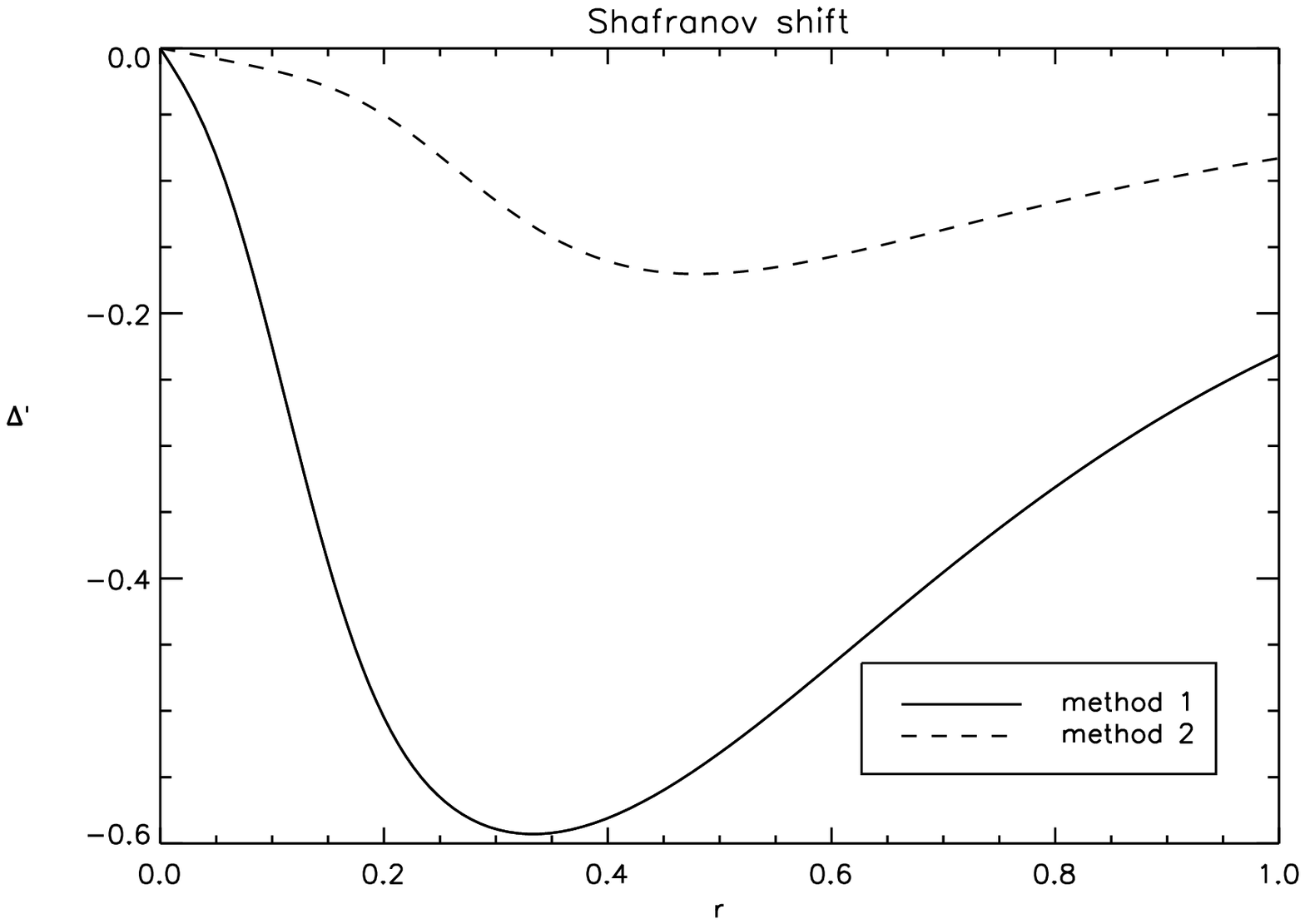}
  \end{tabular}  
  \caption{Left: the safety factor $q$ at the mid-plane for a gravity $g=0.500$.
           Right: the radial derivative of the Shafranov shift $\Delta(r)$ as a function of the radius $r$.}
  \label{fig:cool_strongg_safetyfactor}
\end{figure*}

The last cool solar prominence we discuss is the one for which the dimensionless gravity parameter is $g=1.000$. The gravity is 1000 times stronger than in the first 
discussed equilibrium and 10 times stronger than the last one. The downward shift of the pressure is even stronger as shown in Fig.~\ref{fig:cool_strongg_pressure}. 
For this strong $g$ case, the plasma beta varies from 0.013 at the edge up to 1.107 at the core of the plasma. The latter value is larger than one would expect from 
observations. However, it is straightforward to compute an even more realistic equilibrium by adjusting the coefficient $A_{2}$ of the pressure equation in
Eq.~\eqref{eq:cool_p}. For this equilibrium, the safety factor and the Shafranov shift are shown in Fig.~\ref{fig:cool_strongg_safetyfactor}. Owing to the strong gravity 
the safety factor is very different from its value for the two previous equilibria, particularly in terms of the multiple $q=1$ surfaces. In addition, the plot of the 
Shafranov shift shows a large discrepancy between the two methods, because of the strong gravity. The first method of determining the Shafranov shift is superior in 
terms of accuracy and numerical cost.

\begin{figure*}[ht]
  \centering
  \begin{tabular}{ll}
    \includegraphics[width=0.5\textwidth,trim=3cm 0cm 3cm 0cm]{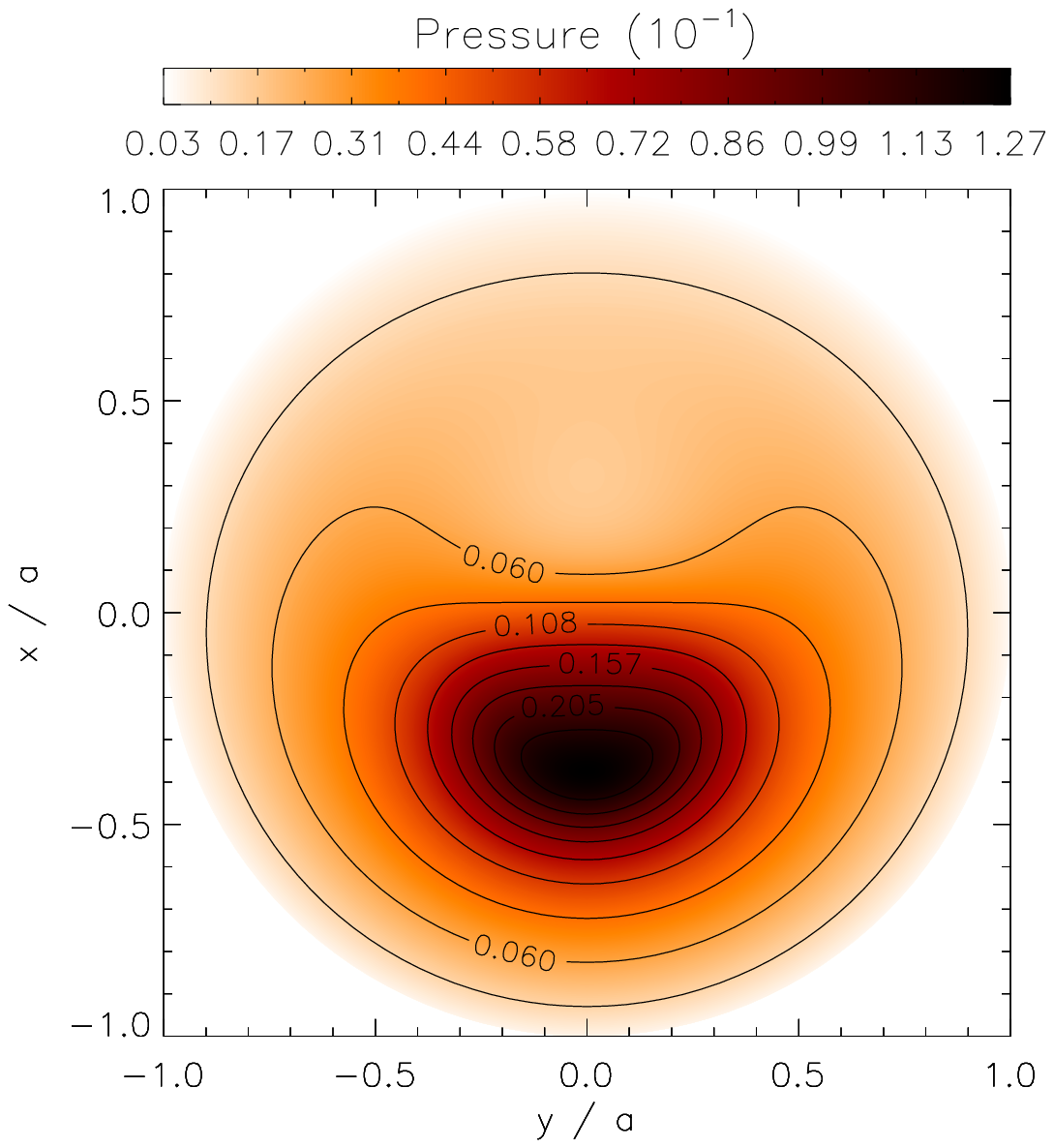} &
    \includegraphics[width=0.5\textwidth,trim=3cm 0cm 3cm 0cm]{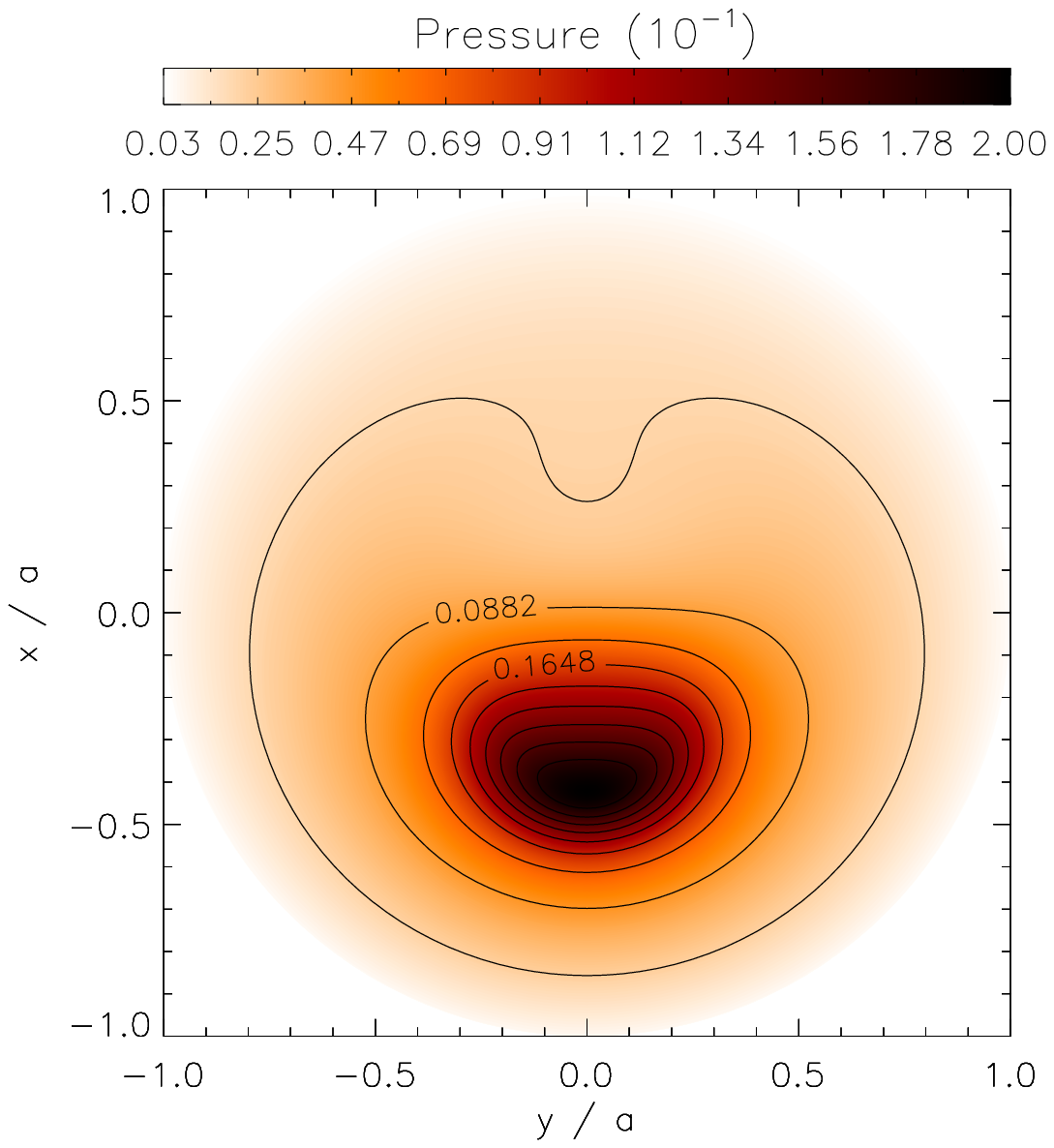}
  \end{tabular}
  \caption{The two-dimensional pressure (red-scale) and plasma beta $\beta=2p/B^{2}$ (contours) profile for a cool solar prominence surrounded by a hot medium with
           a gravity $g=1.000$ for the case that the density (left) or the entropy (right) is a flux function. The solar surface is below the figure. }
  \label{fig:cool_strongg_rho_S}
\end{figure*}

To illustrate the difference between the chosen flux function, we also computed the cases where the density or the entropy is a flux function for gravity
$g=1.000$. For these choices, the pressure and plasma beta are plotted in Fig.~\ref{fig:cool_strongg_rho_S}. When density is a flux function, the plasma beta
varies from 0.011 up to 0.254, and for the entropy case the range is 0.012 up to 0.395. Both plots are clearly distinct from the case where the temperature is 
chosen to be a flux function. From the equilibrium viewpoint, all three are realizable, but high resolution observations will be required to help us to select 
the proper flux function for a particular filament.

\subsection{Multi-layered prominences \label{sec:multi}}
The second class of equilibria, based on the description presented by \cite{Petrie_2007}, allows for multi-layered prominences. This class is specified by the 
following expressions
\begin{align}
  \label{eq:constantT_I}
  I^{2}(\psi) & = A(1 - \tfrac{3}{4} \psi),                                                  \\
  \label{eq:constantT_p}
  p_{0}(\psi) & = AA_{2}(1 - \psi -18\psi^{2} + 56\tfrac{1}{2}\psi^{3}                       \\
              & \quad \phantom{AA_{2}(} - 59\psi^{4} + 20\tfrac{21}{40}\psi^{5}),  \nonumber \\
  \label{eq:constantT_T}
  T_{0}(\psi) & = A_{3},                                                                     \\
  \label{eq:constantT_vz}
  v_{z}(\psi) & = 0,
\end{align}
where $A_{2}=0.1$ and $A_{3}=1$. In addition, the overall amplitude A is computed by FINESSE. The gravity $g$ is varied from $0.001$ up to $5.000$. As for the
previous class of cool prominences, we assume that the boundary is circular and that the temperature is a flux function. The function $T_{0}$ has the same meaning as in 
the previous subsection, and represents the (quasi) temperature of the chosen flux function. As for the previous class, the numerical computations are 
done for 101 and 129 points in the radial and poloidal direction, respectively.

\begin{figure*}[ht]
  \centering
    \includegraphics[width=0.6\textwidth,trim=3cm 0cm 3cm 0cm]{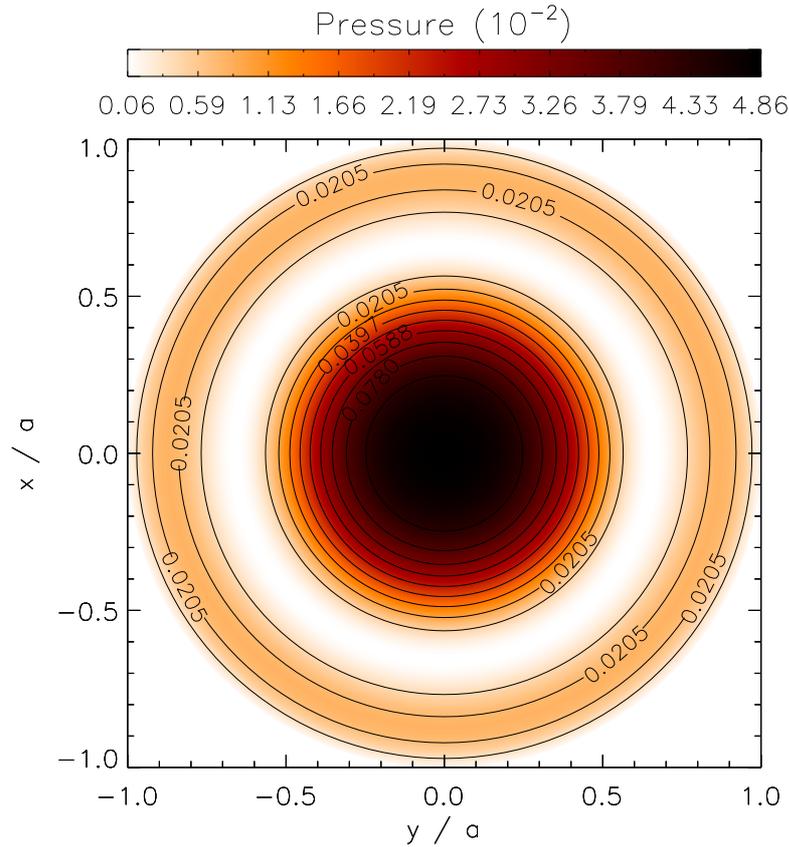}
  \caption{The two-dimensional pressure (red-scale) and plasma beta $\beta=2p/B^{2}$ (contours) profile for a double-layered solar prominence with a gravity $g=0.001$. 
           The solar surface is below the figure.}
  \label{fig:constantT_smallg_pressure}
\end{figure*}

\begin{figure*}[ht]
  \centering
  \begin{tabular}{ll}
    \includegraphics[width=0.5\textwidth]{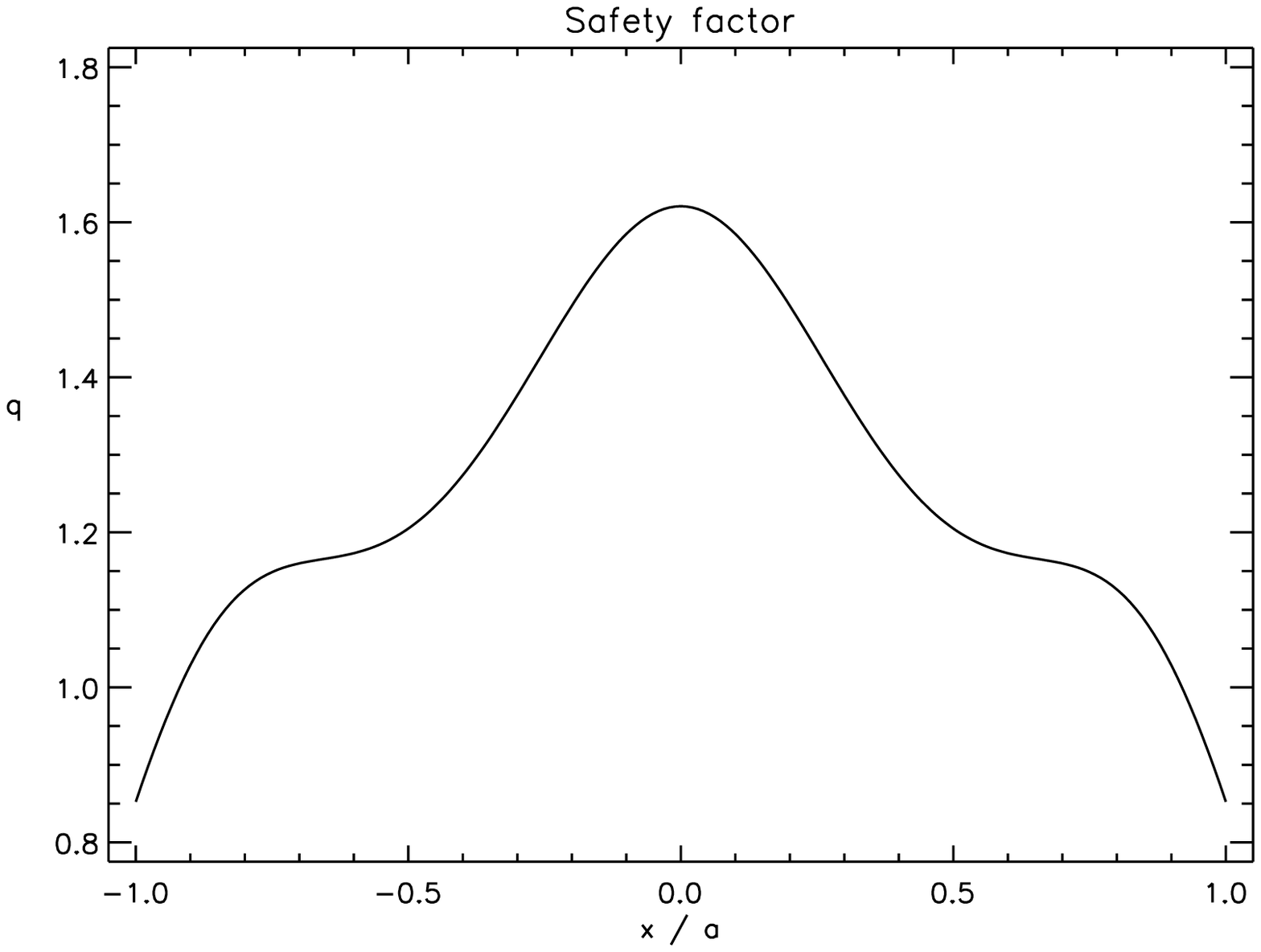} &
    \includegraphics[width=0.5\textwidth]{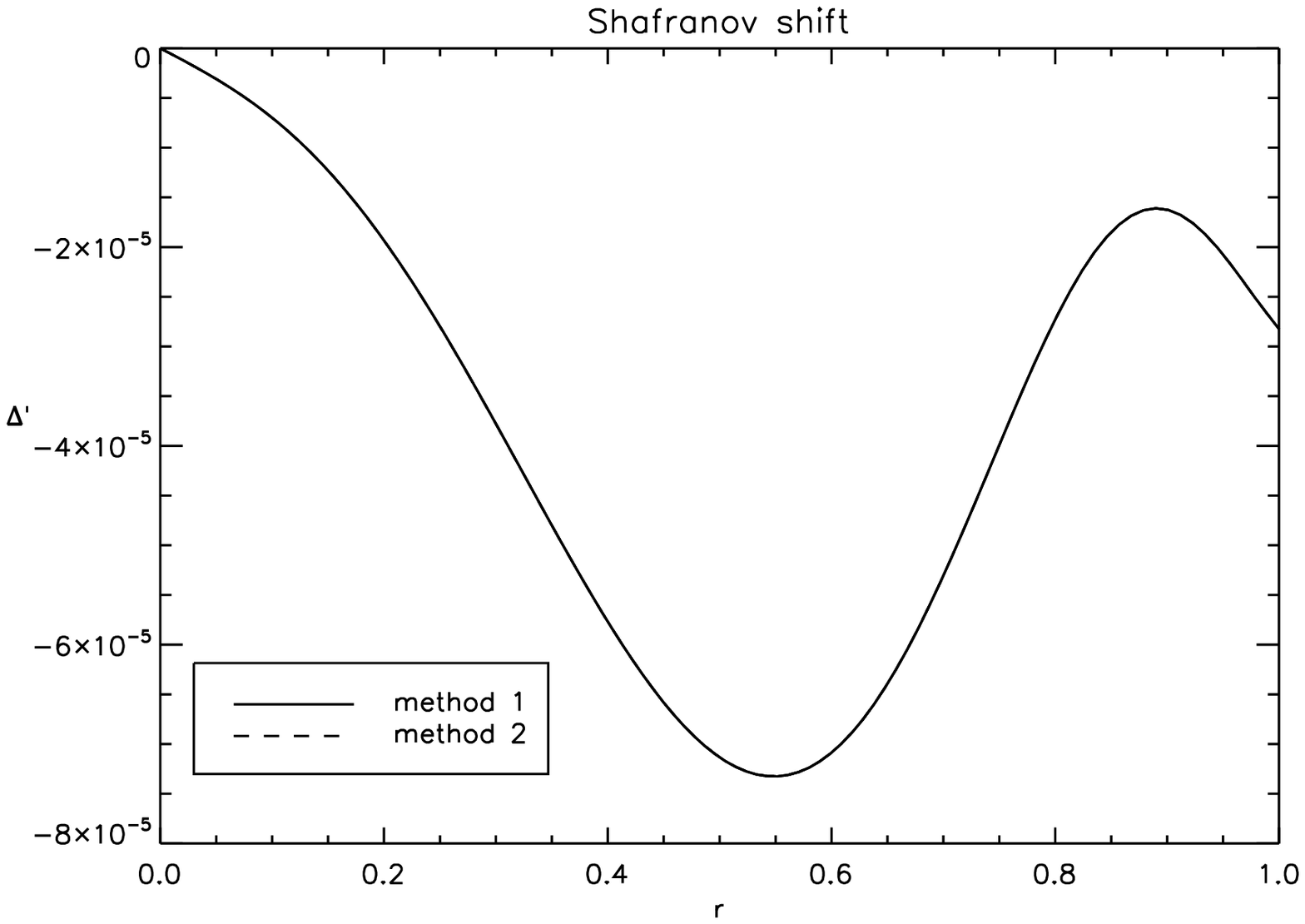}
  \end{tabular}  
  \caption{Left: the safety factor $q$ at the mid-plane for a gravity $g=0.001$.
           Right: the radial derivative of the Shafranov shift $\Delta(r)$ as a function of the radius $r$.}
  \label{fig:constantT_smallg_safetyfactor}
\end{figure*}

Starting with small gravity $g=0.001$, Fig.~\ref{fig:constantT_smallg_pressure} shows the two-dimensional pressure and the plasma beta $\beta = 2p/B^{2}$. The latter 
quantity varies from 0.001 at the edge to 0.097 at the center of the prominence. Once more, we emphasize that
for these plasma beta values the magnetic field as well as the pressure play an important role in the stability analysis of our accompanying paper. The pressure
shows a ring structure. By increasing the parameter $g$, we show below that one can create a double-layered prominence. The safety factor $q$
and the radial derivative of the Shafranov shift are plotted in Fig.~\ref{fig:constantT_smallg_safetyfactor}. For small $g$, the safety factor $q$ has two flat regions 
that correspond to the ring-like cavity of low pressure. Furthermore, note that the $q=1$ and $q=3/2$ surfaces both exist for this equilibrium. As for cool 
prominences, this will mean that gaps or instabilities may occur in the MHD continuous spectrum. The figure also shows that the two methods for computing the Shafranov 
shift, as discussed in the previous subsection, show excellent agreement. 

\begin{figure*}[ht]
  \centering
  \includegraphics[width=0.6\textwidth,trim=3cm 0cm 3cm 0cm]{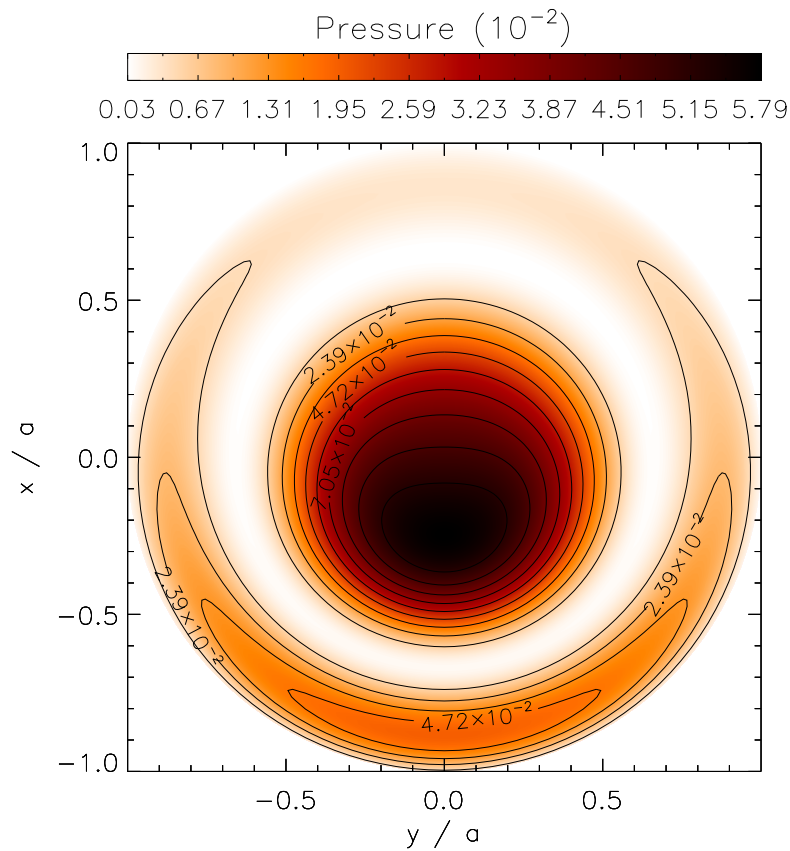}
  \caption{The two-dimensional pressure (red-scale) and plasma beta $\beta=2p/B^{2}$ (contours) profile for a double-layered solar prominence with a gravity $g=1.000$. 
           The solar surface is below the figure.}
  \label{fig:constantT_intermediateg_pressure}
\end{figure*}

\begin{figure*}[ht]
  \centering
  \begin{tabular}{ll}
    \includegraphics[width=0.5\textwidth]{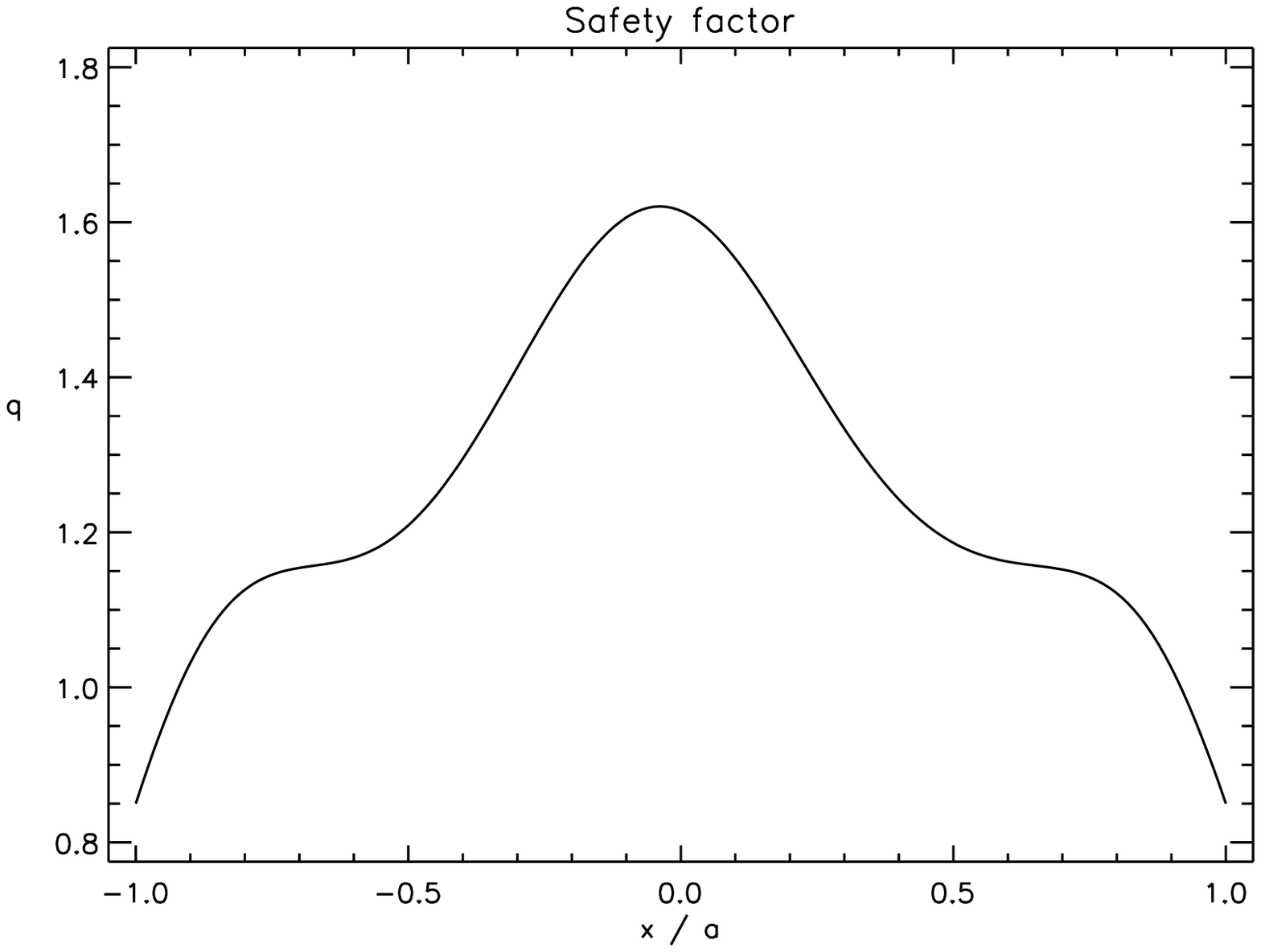} &
    \includegraphics[width=0.5\textwidth]{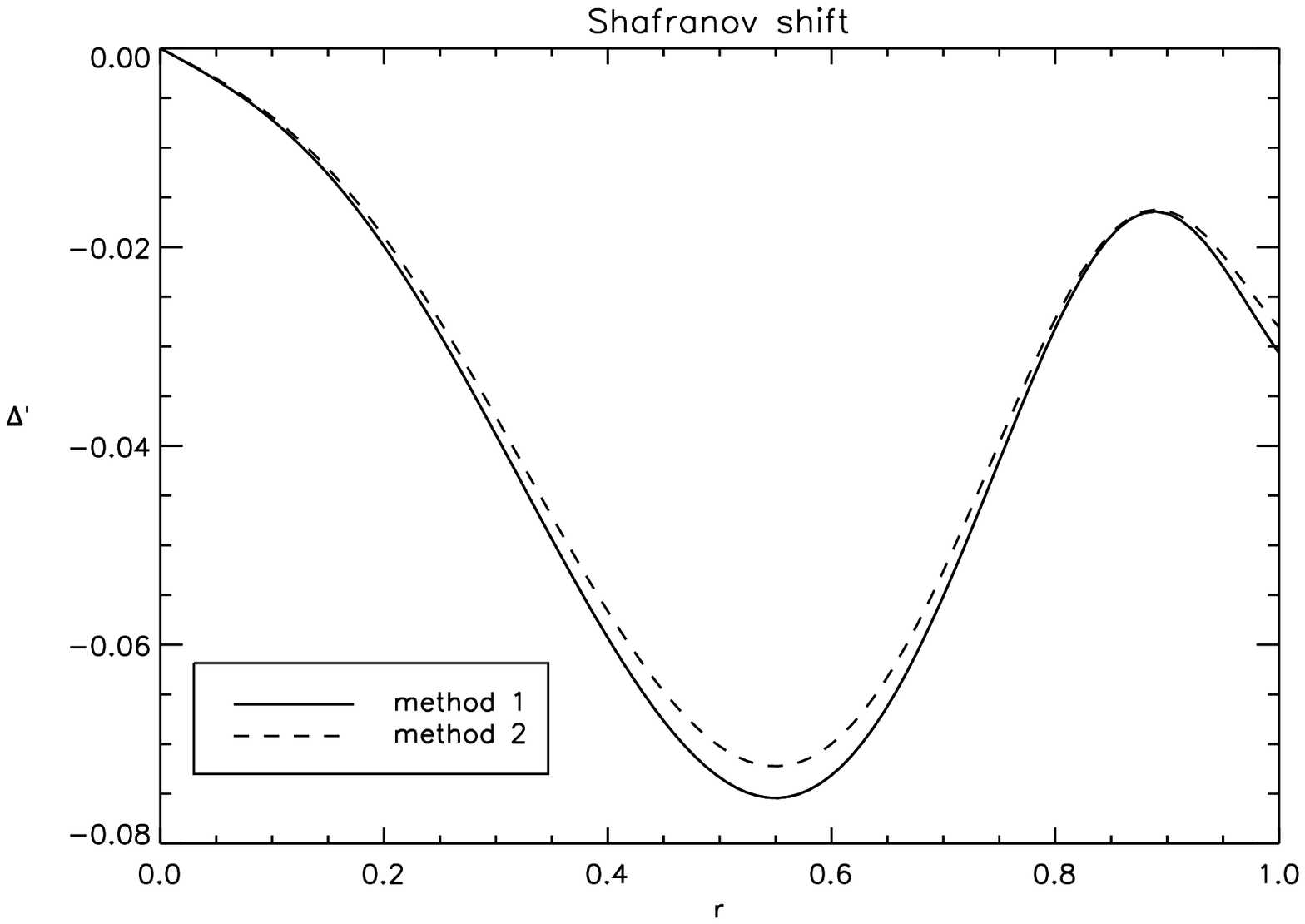}
  \end{tabular}  
  \caption{Left: the safety factor $q$ at the mid-plane for a gravity $g=1.000$.
           Right: the radial derivative of the Shafranov shift $\Delta(r)$ as a function of the radius $r$.}
  \label{fig:constantT_intermediateg_safetyfactor}
\end{figure*}

For the next equilibrium, we have set the gravity parameter $g=1.000$, which is 1000 times stronger than in the case above. This example shows the onset of the 
double-layered pressure profile as can be seen in Fig.~\ref{fig:constantT_intermediateg_pressure}. If one closely looks at the plot, one still sees the ring structure 
but because of the stronger gravity compared to the small gravity $g=0.001$ case this structure has almost disappeared. The plasma beta ranges from 0.00069 in the low 
pressure regions up to 0.117 in the high pressure regions. For this equilibrium, we also computed the safety factor $q$ and the Shafranov shift. Both quantities are 
presented in Fig.~\ref{fig:constantT_intermediateg_safetyfactor}. A comparison between the safety factor for this equilibrium and that for the small gravity 
$g=0.001$ case finds hardly any differences. The only real difference is the maximum of the $q$ profile, which is shifted slightly downwards because of the stronger 
gravity. The plot of the Shafranov shift shows that a small difference between both means of determining the shift is clearly evident. The second way of applying
the Shafranov equation given in Eq.~\eqref{eq:shafranovshift} again underestimates the radial derivative.

\begin{figure*}[ht]
  \centering
  \includegraphics[width=0.6\textwidth,trim=3cm 0cm 3cm 0cm]{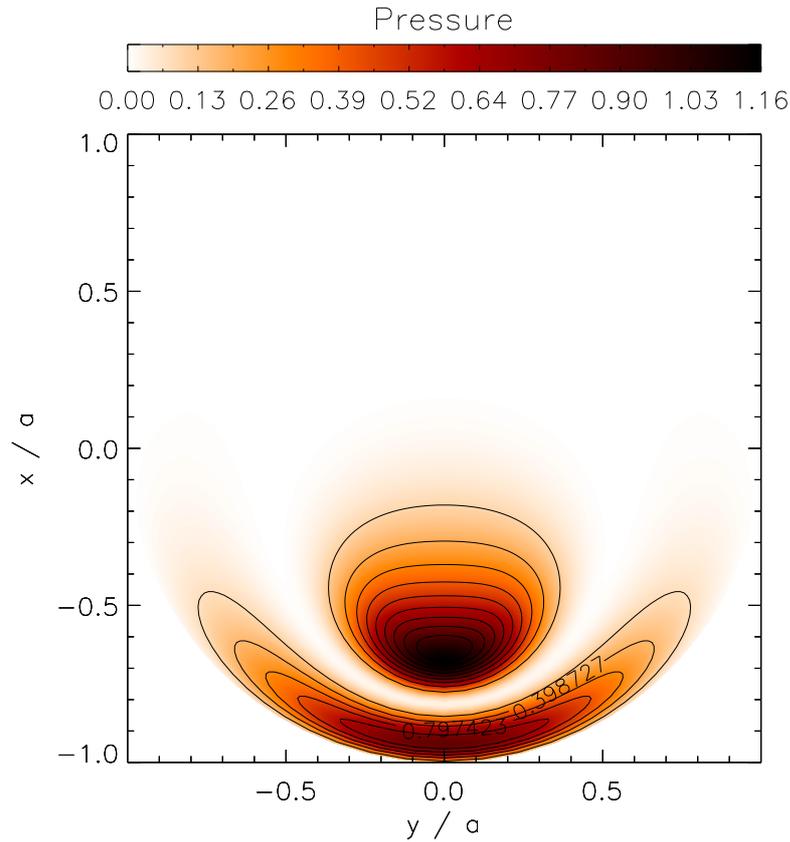}
  \caption{The two-dimensional pressure (red-scale) and plasma beta $\beta=2p/B^{2}$ (contours) profile for a double-layered solar prominence with a gravity $g=5.000$. 
           The solar surface is below the figure.}
  \label{fig:constantT_strongg_pressure}
\end{figure*}

\begin{figure*}[ht]
  \centering
  \begin{tabular}{ll}
    \includegraphics[width=0.5\textwidth]{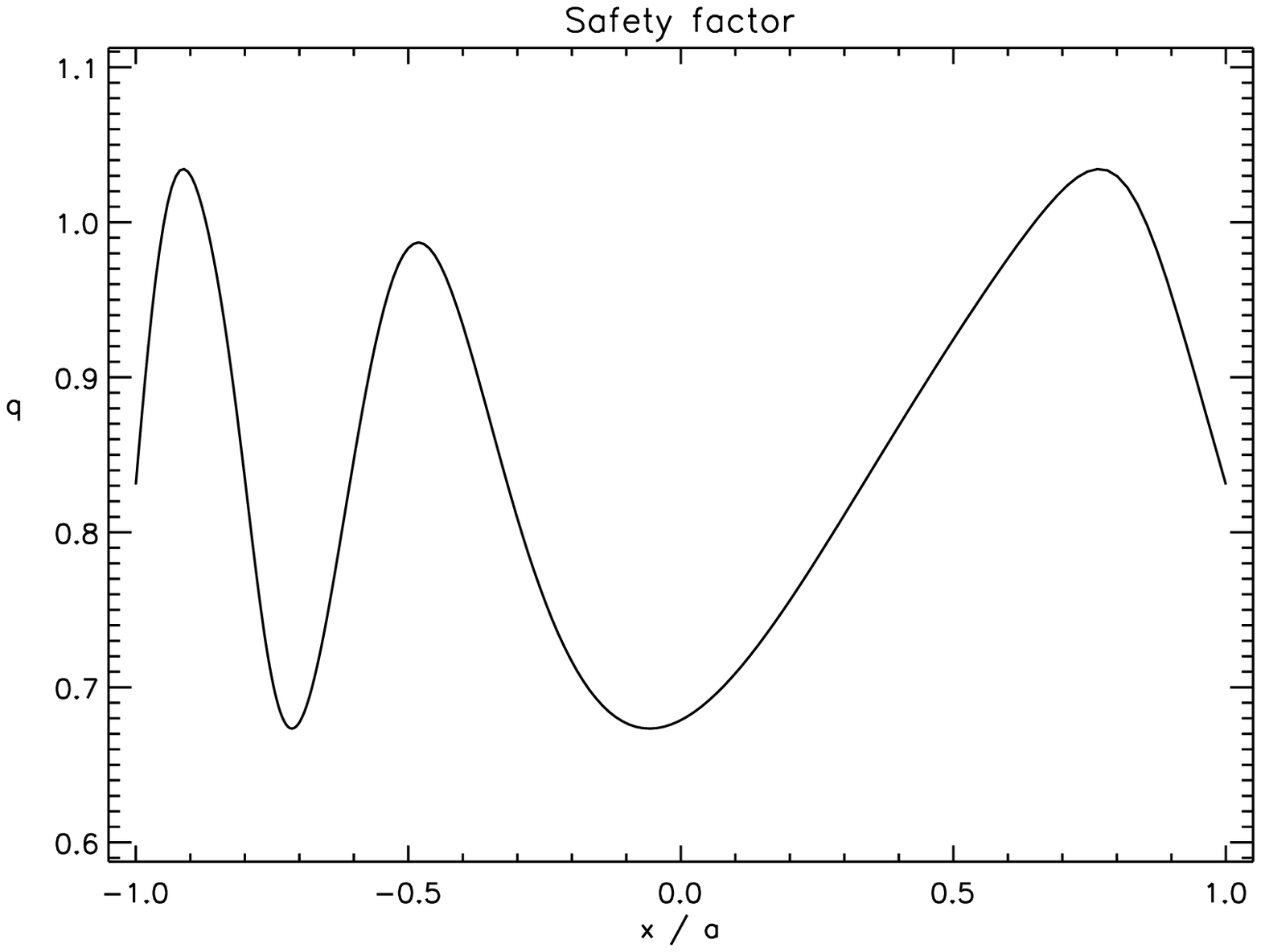} &
    \includegraphics[width=0.5\textwidth]{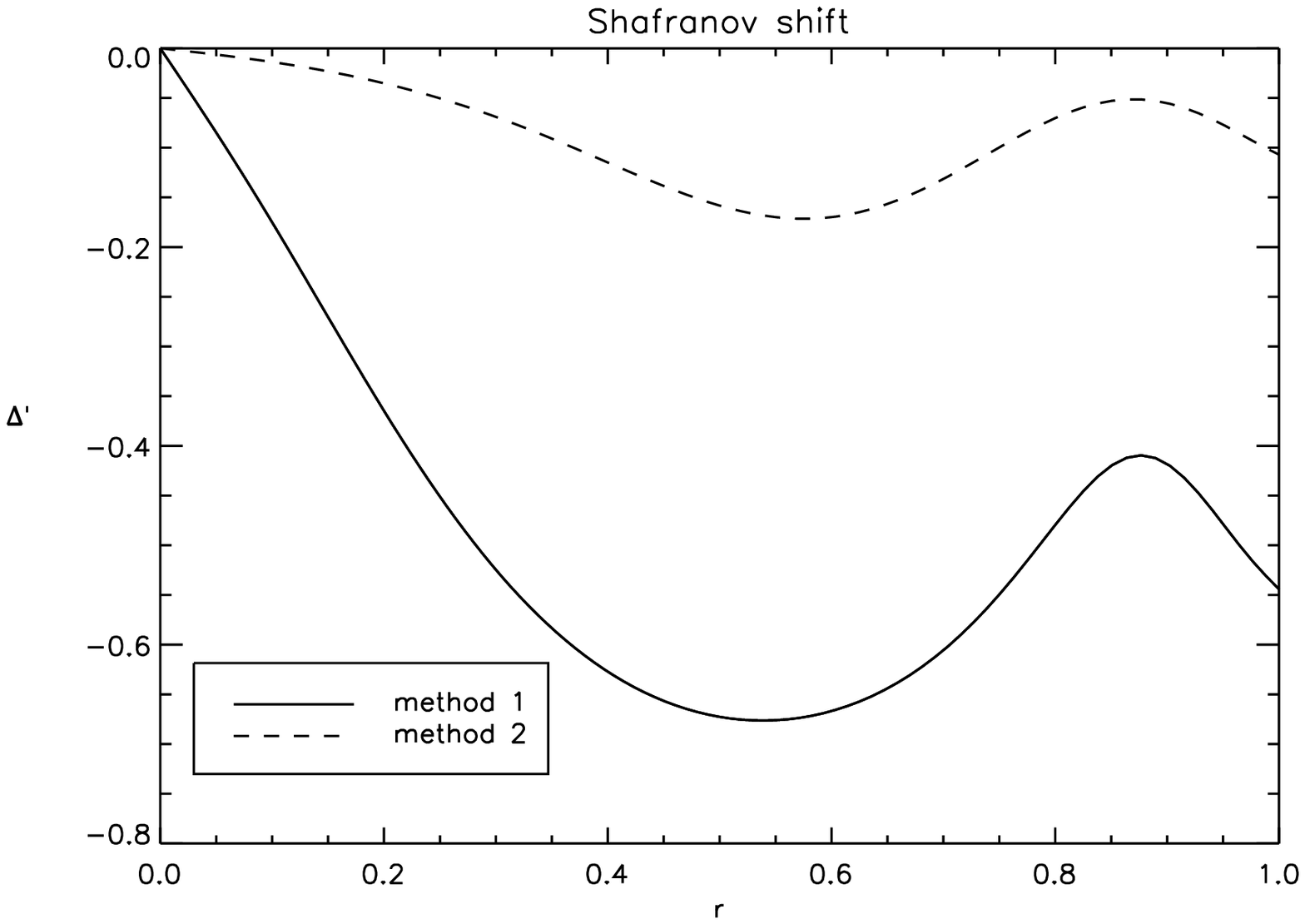}
  \end{tabular}  
  \caption{Left: the safety factor $q$ at the mid-plane for a gravity $g=5.000$.
           Right: the radial derivative of the Shafranov shift $\Delta(r)$ as a function of the radius $r$.}
  \label{fig:constantT_strongg_safetyfactor}
\end{figure*}

\begin{figure*}[ht]
  \centering
  \includegraphics[width=\textwidth]{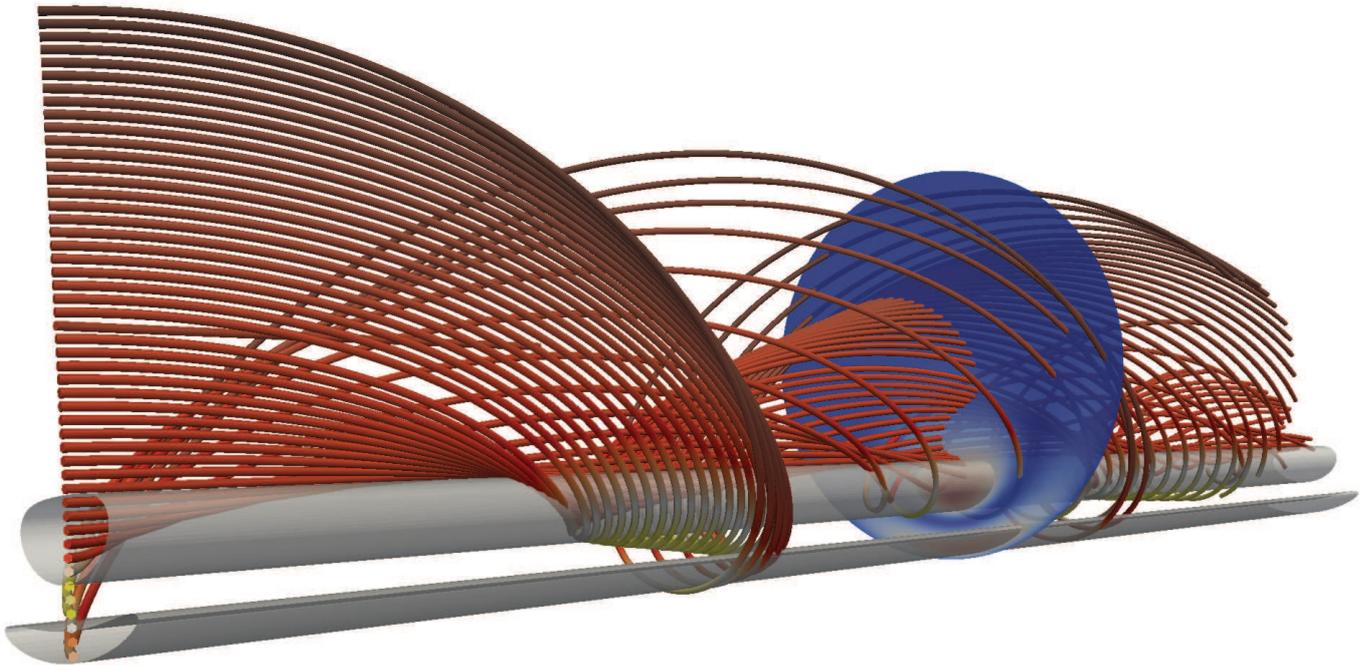}
  \caption{A three-dimensional impression of the double-layered solar prominence for gravity parameter $g=5.000$. Shown are an isosurface of density, at a value 
           showing the layering of the condensation, with the cross-sectional view shown midway this 3D impression. The magnetic field structure is visualized by 
           drawing selected field-lines, demonstrating the varying twist of the helical configuration. The field lines are colored by the magnetic field strength. 
           The solar surface is below the figure.}
  \label{fig:prominence3d}
\end{figure*}

The increase in the gravity parameter to $g=5.000$ creates a clear double-layered pressure structure as can be seen in Fig.~\ref{fig:constantT_strongg_pressure}. 
A similar structure is found for the density. The plasma beta varies from 0.000031 up to 1.994. The latter value is higher than we expect based on
observations. As mentioned before, it is straightforward to compute a more realistic value by changing the coefficient $A_{2}$ of the profile pressure 
equation in Eq.~\eqref{eq:constantT_p}. The safety factor of this double-layered prominence, shown in Fig.~\ref{fig:constantT_strongg_safetyfactor}, varies widely. 
It no longer contains a $q=3/2$ surface as in the previous two cases of this equilibrium class. Furthermore, we note that near the magnetic axis 
(around $x=-0.5$) it almost creates new $q=1$ surfaces. The magnetic axis is the location where the poloidal magnetic field is zero. The Shafranov shift is also shown in 
Fig.~\ref{fig:constantT_strongg_safetyfactor}. As expected, the two methods for determining the shift no longer agree. As in the previously discussed case, the 
Shafranov shift equation in Eq.~\eqref{eq:shafranovshift} underestimates the radial derivative. Figure~\ref{fig:prominence3d} presents this prominence configuration 
in 3D, illustrating the varying twist of the magnetic field lines. In our accompanying paper, we will analyze the way in which this more realistic field configuration 
modifies the MHD eigenspectrum of the flux-surface localized continuum modes.


\section{Conclusions \label{sec:conclusions}}
We have presented the equations describing translational symmetric solar prominence equilibria. These equations have been specialized for three choices
of the freedom in flux function in which the temperature, density, or entropy can be flux-dependent. For small gravity and circular cross-section, we have expanded these 
equations to derive the equation for the equilibrium of a cylindrical plasma and the equation for the Shafranov shift.

The MHD equilibrium code FINESSE has been used to compute accurate prominence equilibria. A comparison between the numerically generated equilibrium
and the Dungey solution shows fourth order convergence, and moderate resolution is needed to obtain an accurate solution.

We have considered two equilibrium classes, cool prominences surrounded by a hot medium and double-layered prominences, for different values of the dimensionless gravity 
parameter (in a range corresponding to solar filament cases) with plasma beta of the order 0.1. For both classes, a comparison between the numerical solution and the 
derived equation for the Shafranov shift has been made. This comparison shows excellent agreement for small gravity, while for strong gravity only method 1 is able to
quantify the shift.

For cool prominences, the results show that by increasing the gravity, the location of the maximum pressure shifts in the downwards direction. This downward shift is 
largest when the temperature is a flux function. Furthermore, in a strong gravitational potential the three choices of the chosen flux function 
show large deviations from each other in, for example, the pressure and density. This means that observations could indicate what the most suitable choice of 
the flux function should be.

The results of the double-layered prominences have revealed that a double-layered structure in pressure and density can be created in actual filament configurations. 
The relative strength of the gravitational potential must be sufficient: if the potential is weak, a ring structure appears, where a cavity surrounds the prominence. 
As for cool prominences, the location of the maximum pressure and density is shifted increasingly downwards if the gravity importance is increased.

In our accompanying paper \citep{Blokland_2011B}, the stability properties of these equilibria will be analyzed, with special attention to the continuous MHD spectrum. 
Owing the presence of gravity, gaps or even instabilities may appear in this continuous spectrum. Furthermore, inside these gaps new global modes may occur, 
which provide us with important information about the internal structure of the prominence. Before investigating the possible appearance of global modes, 
a detailed analysis of the continuous spectrum will be required.

\begin{acknowledgements}
This work was carried out within the framework of the European Fusion Programme, supported by the European Communities under contract of the Association EURATOM/FOM. 
Views and opinions expressed herein do not necessarily reflect those of the European Commission. RK acknowledges financial support by project GOA/2009/009 (K.U.Leuven). 
The research leading to these results has received funding from the European Commission's Seventh Framework Programme (FP7/2007-2013) under the grant agreement SWIFF 
(project nr 263340, www.swiff.eu).
\end{acknowledgements}

\bibliographystyle{aa}
\bibliography{references}

\end{document}